\documentstyle[aps,epsf]{revtex}
\newcommand{\be}{\begin{equation}}
\newcommand{\ee}{\end{equation}}
\newcommand{\bea}{\begin{eqnarray}}
\newcommand{\eea}{\end{eqnarray}}

\begin{document}
\title{
Instabilities and Spatio-Temporal Chaos in Hexagon Patterns 
with Rotation}
\author{Filip Sain\footnote{Corresponding author. Tel.: (847) 
491-3345,
Fax.: (847) 491-2178, email: fsain@nwu.edu.}
 and Hermann Riecke}
\address{Department of Engineering Science and Applied Mathematics\\
Northwestern University, 2145 Sheridan Road, Evanston, IL 60208, USA} 
\maketitle
%%%%%%%%%%%%%%%%%%%%%%%%%%%%%%%%%%%%%%%%%%%%%%%%%%%%%%%%%%%%%%%%%%%%%%%%%%%%%%%%

\begin{abstract}

The dynamics of hexagon patterns in rotating systems are investigated
within the framework of modified Swift-Hohenberg equations that can
be considered as simple models for rotating convection with broken 
up-down symmetry, e.g.  non-Boussinesq Rayleigh-B\'{e}nard or 
Marangoni convection.  In the weakly nonlinear
regime a linear stability analysis of the hexagons reveals long- and short-wave
instabilities, which can be steady or oscillatory. The oscillatory short-wave 
instabilities can lead to stable hexagon patterns that are periodically 
modulated in space and time, or to a state of spatio-temporal chaos with
a Fourier spectrum that precesses on average in time. The chaotic state 
can exhibit bistability with the steady hexagon pattern. There exist regimes
in which the steady hexagon patterns are unstable at all wavenumbers.



\end{abstract}

%%%%%%%%%%%%%%%%%%%%%%%%%%%%%%%%%%%%%%%%%%%%%%%%%%%%%%%%%%%%%%%%%%%%%%%%%%%%%%

\section{Introduction}

Patterns arising from hydrodynamic instabilities have become a 
paradigm for the investigation
of dynamical systems with many degrees of freedom. In particular 
Rayleigh-B\'enard convection 
has been studied experimentally and theoretically in great detail. In 
recent years the
focus of investigation has turned to patterns that are spatially and 
possibly also
temporally complex. Various types of such states of spatio-temporal 
chaos have been
identified. Great interest has, for example, found the spiral-defect 
chaos that is observed in convection of
fluids with not too large Prandtl number \cite{MoBo93,DePe94a}.
It is characterized by the appearance and disappearance of spirals in 
the convection patterns that are separated by domains of straight or 
slightly curved convection rolls. 
Spiral-defect chaos has not been found directly at onset of 
convection. 
Thus, it is not accessible by weakly nonlinear theory.
From a theoretical point of view another type of spatio-temporal chaos
has therefore been very attractive since it arises directly at onset. 
It is observed in rotating convection and is due to the 
K\"{u}ppers-Lortz 
instability through which all straight-roll 
states become unstable to rolls of a different orientation if the
rotation rate is sufficiently large \cite{KuLo69}. 
In a simple three-mode model for rolls rotated
by $120^o$ with respect to each other this instability leads 
to a structurally stable heteroclinic
orbit connecting the three sets of rolls 
\cite{BuHe80,MaLe75}. 
In large aspect-ratio experiments \cite{ZhEc91,HuEc95,HuEc97} 
and numerical simulations of Ginzburg-Landau equations \cite{TuCr92} 
and of order-parameter models \cite{CrMe94,XiGu94} 
the switching between the different orientations looses coherence
across the system and domains
of almost straight rolls of different orientations form that persistently
invade each other in an irregular manner. 

The character of a spatio-temporally chaotic state is expected to 
depend strongly on the planform
of the underlying pattern. Thus, spatio-temporal chaos arising from 
instabilities of 
stripes (or rolls) will differ from that of square or hexagonal 
patterns. For instance,
in hexagon (or square) patterns the bending of rolls that is widely 
observed in
Rayleigh-B\'enard convection is strongly suppressed. Instead, the 
appearance of disorder in 
a hexagonal pattern might resemble more the melting of a `crystal'.
There have been only a few studies of transitions to spatio-temporal
chaos from square or hexagon patterns. Experimental investigations have been made
on parametrically driven surface waves
\cite{TuRa89,KuGo96a} and on a chemical system \cite{OuSw91}.

In the present work we consider hexagon patterns in systems with broken
up-down symmetry 
and study the effect of a breaking of the chiral symmetry on their 
stability and dynamics.
It is motivated by the complex dynamics ensuing from the 
K\"uppers-Lortz instability of 
convection rolls, where the chiral symmetry is broken by a global 
rotation of the system. 
From a general weakly nonlinear analysis of the three modes making
up the hexagon patterns 
it is known that a breaking of the up-down symmetry transforms
the structurally stable heteroclinic cycle that 
results from the K\"uppers-Lortz instability within the three-mode 
model into a periodic orbit of oscillating hexagons \cite{Sw84,So85,MiPe92}.
This periodic orbit arises from a Hopf bifurcation, which replaces 
the steady instability 
of the hexagonal pattern that leads to rolls \cite{Sw84,So85}. 

We are interested in the question
which additional instabilities can arise if one goes beyond the 
three-mode model and 
allows arbitrary perturbation modes. Of particular interest are 
situations in which 
such perturbations lead to persistent dynamics. One possible approach 
is to extend the 
three-mode amplitude equations to three coupled Ginzburg-Landau 
equations that also allow for
a slow spatial variation of the amplitudes \cite{EcRi99}. The slowness of the
space dependence implies, 
however, that all perturbations to the hexagons are restricted to wavevectors
close to those of the hexagon modes. In particular, only modes making 
a small angle with 
respect to the hexagon modes can be treated. Furthermore,
the Ginzburg-Landau
equations cannot be expected to be suitable for general numerical 
simulations of
unstable hexagon patterns in large systems since the choice of the 
three 
carrier modes making up the hexagons strongly breaks the isotropy of 
the system. Only in the unlikely case that
the perturbation wavevectors remain close to the wavevectors of the hexagon modes
for all times
would the Ginzburg-Landau equations be adequate.  

 The goal of our investigation is to explore what type of 
instabilities can appear 
due to the breaking of the chiral symmetry and whether they can lead 
to persistent 
regular or irregular dynamics. It is therefore important 
to allow arbitrary perturbation modes and to preserve 
the isotropy of the system. Thus, we investigate a simple model of 
the 
Swift-Hohenberg type \cite{SwHo77,NeFr93,BeFr98}. Such models have
been widely used 
to address general questions regarding the dynamics of patterns 
(e.g. \cite{MiPe92,NeFr93,XiGu94,KuHe96,SaBr97a,CrRi99}).
If the patterns arise from a long-wave
instability equations of the Swift-Hohenberg type can be derived systematically in a
long-wave analysis.
This has been done, for instance, for buoyancy-driven convection with
asymmetric boundary conditions \cite{Co98}
and for Marangoni convection \cite{Kn90,MaSaunpub}. In both cases the onset of
convection
occurs at long wavelengths if the boundaries are poor conductors.
If the wavenumber at onset
is finite the Swift-Hohenberg cannot be derived rigorously from the basic equations
(e.g. Navier-Stokes equations), but they can be obtained
as approximate order-parameter models by expanding the nonlinear 
interaction terms (e.g. \cite{NeFr93,BeFr98}).

We perform a general linear stability analysis of the weakly 
nonlinear hexagon pattern 
and pay particular attention to destabilizing modes that are rotated 
with respect to the hexagon pattern.
In the absence of rotation such general stability analyses of 
hexagonal 
patterns have been performed for the
full fluid equations for Marangoni convection \cite{Be93} as well as 
within 
coupled Ginzburg-Landau equations \cite{SuTs94}. Some results are 
also available for
weakly nonlinear Marangoni convection in the presence of rotation 
\cite{Ri94}.
We complement the stability analysis with direct simulations of the 
order-parameter
equation to investigate the nonlinear behavior resulting from the 
instabilities. 
We find that the hexagonal pattern can become unstable at all 
wavelengths, resulting
in spatio-temporally chaotic states. 
Moreover, even for parameters for which stable hexagon patterns 
exist, 
persistent periodic and irregular dynamics are found, implying 
bistability of ordered and
disordered states.

The paper is organized as follows. 
In Section II we introduce two order-parameter models and set up the
linear 
stability analysis of the weakly nonlinear hexagons. 
In Section III the results of that analysis are presented. 
We discuss the different types of instabilities and the resulting 
stability 
regions of hexagonal patterns. 
In Section IV numerical simulations of the order-parameter equation 
are used 
to confirm the stability analysis and to investigate the nonlinear 
evolution 
ensuing from it. 
Conclusions are found in Section V.

%%%%%%%%%%%%%%%%%%%%%%%%%%%%%%%%%%%%%%%%%%%%%%%%%%%%%%%%%%%%%%%%%%%%%%%%%%%%%%%%%%%%%%%%

\section{The Model and Linear Stability Analysis}

Spatially periodic hexagon patterns with small amplitude arising
from a weakly transcritical bifurcation can be
described systematically 
by coupled equations for the complex amplitudes $A$, $B$, and $C$
 of the three stripe (roll) components 
\begin{equation} \frac{\partial A}{\partial \tau_{2}}=R_{2} 
A+2\,\zeta 
B^{*}C^{*}+\left(f_{1}{|A|}^{2}+f_{2}{|B|}^{2}+f_{3}{|C|}^{2}\right)A,
\label{eq:ampeq}
\ee 
where $\tau_{2}$ is a slow time, $R_{2}$ a reduced control parameter
and the difference between the cubic coupling coefficients $f_{2}$ 
and $f_{3}$ is proportional to the breaking of the chiral symmetry. 
The equations for $B$ and $C$ follow by cyclic permutation.
To investigate the stability of these hexagons with respect to 
spatially varying perturbations the amplitudes could be allowed to 
be space-dependent which would lead to the inclusion of spatial 
derivatives in (\ref{eq:ampeq}). The amplitude description breaks,
however, the isotropy of 
the system and requires that the 
space dependence of the amplitudes be slow. This implies that only 
perturbations with wavevectors close to those making up the hexagon can be 
considered. In this study we are interested in particular 
in perturbations that are rotated with respect to the hexagons by an 
arbitrary amount and in dynamical, disordered states that may
exhibit an essentially isotropic wavevector spectrum. Therefore the
isotropy of the system needs to be preserved and 
Ginzburg-Landau-type amplitude equations are not sufficient to 
investigate the questions we are interested in. 

For a first exploration of the impact of a chiral symmetry
breaking on side-band perturbations of hexagons and their nonlinear 
evolution it is appropriate to preserve the isotropy of the system
by considering a simple model equation rather than the full 
hydrodynamical problem. Detailed investigations of specific physical systems
can then be performed as a second step guided by the results of the model equations.
 
In a number of studies Swift-Hohenberg-type models have provided 
valuable insight into the dynamics of patterns
(e.g. \cite{MiPe92,NeFr93,XiGu94,KuHe96,SaBr97a,CrRi99}). We
consider here a minimal model of that type for an order
parameter $\psi(x,y,t)$. In order to 
obtain stable hexagons in the absence of rotation we introduce a
quadratic term $\alpha \psi^{2}$ that breaks the up-down symmetry
$\psi \rightarrow -\psi$. The cubic scalar term $\psi^{3}$ of the
original Swift-Hohenberg equation
provides nonlinear saturation. To break the chiral symmetry
nonlinear gradient terms have to be introduced. At leading order in 
the amplitude and the gradients this introduces an additional quadratic term and we are
led to consider the model
\begin{eqnarray} \partial_{t}\psi& = &R\psi-(\nabla^{2}+1)^{2}\psi-\psi^{3}
       +\alpha \psi^{2} +\gamma\;\hat{e}_{z} \cdot (\nabla
\psi \times \nabla (\nabla^{2} \psi)), \label{eq:mshe}
 \end{eqnarray}  
where $\hat{e}_{z}$ is the unit vector
perpendicular to the $(x,y)$-plane. In these equations the bifurcation 
parameter is $R$ and the wavenumber of the bifurcating mode has been 
scaled to 1. The strength of the chiral symmetry-breaking is measured 
by $\gamma$.

It should be noted that the quadratic gradient term in (\ref{eq:mshe})
has the somewhat non-generic feature that it generates harmonics only from the 
interaction of modes with wavevectors of different magnitude. As a consequence,
in the
small-amplitude regime (\ref{eq:ampeq}), for which $\alpha$ has to be
assumed to be small, it does not modify the cubic
coefficients $f_i$ in (\ref{eq:ampeq}) and  has therefore to this order no effect
on strictly periodic hexagon patterns. In particular, the instability of hexagons that
occurs with increasing $R$ is not oscillatory and does not lead to the periodic
orbit of oscillating hexagons identified in \cite{Sw84,So85} (cf. Fig.\ref{fig:bifdiags}).
 Instead, the instability remains steady and leads directly to the
roll solution as is the case in the
absence of rotation.
In this paper we focus on the stability of 
the steady hexagons with respect to side-band perturbations 
{\it below} that instability. Thus, the existence of the secondary branch of
oscillating hexagons is not of central importance.
We present most results for the minimal model (\ref{eq:mshe}).

The non-genericity of the quadratic gradient term in (\ref{eq:mshe}) can be addressed
by introducing additional terms.
One possibility is to introduce a quadratic term,
e.g. $\beta (\nabla \psi)^2$, that allows $\alpha$ to be of
$O(1)$ as long as $\alpha$ and 
$\beta$ together lead to $\zeta \ll1$ in (\ref{eq:ampeq}).
Alternatively, a cubic term that breaks the chiral symmetry can be 
introduced, e.g. $g_{2}\;\hat{e}_{z} \cdot (\nabla \times (\nabla \psi)^{2} \nabla \psi)$.
We thus have the extended model
\begin{eqnarray} \partial_{t}\psi& = &R\psi-(\nabla^{2}+1)^{2}\psi-\psi^{3}
       +\alpha \psi^{2} +\beta (\nabla \psi)^2 + \gamma\;\hat{e}_{z} \cdot (\nabla
\psi \times \nabla (\nabla^{2} \psi)) 
+g_{2}\hat{e}_{z} \cdot ( \nabla \times (\nabla \psi)^{2} \nabla 
\psi), \label{eq:mshe2}
 \end{eqnarray}  
whose results we discuss briefly.
 
While we are using (\ref{eq:mshe}) and (\ref{eq:mshe2})
as simple model equations,
equations of that type can be obtained from the basic physical equation 
(e.g. the fluid equations) by expanding the interaction term in Fourier space
\cite{NeFr93,Ha83a,BeFr99}. In general, one then obtains additional
nonlinear gradient terms. Similar equations can be derived  
systematically as long-wave equations (e.g.
for buoyancy-driven convection with asymmetric boundary conditions 
\cite{Co98} or for surface-tension driven convection without \cite{Kn90} and
with rotation \cite{MaSaunpub}).

We perform a weakly nonlinear analysis of (\ref{eq:mshe}) and (\ref{eq:mshe2}),
and start with a small amplitude
expansion of $\psi$ on a hexagonal lattice,
\begin{eqnarray} 
\psi& = &\hspace{.20in}\epsilon 
\left\{Ae^{ikx}+Be^{ik(-x/2+y\sqrt{3}/2)}+Ce^{ik(-x/2-y\sqrt{3}/2)}\right\}\label{eq:psihex}\\                      
&   &\mbox{}+\epsilon^{2} 
\left\{D_{000}+D_{200}e^{i2kx}+D_{020}e^{ik(-x+\sqrt{3}y)}+
D_{002}e^{ik(-x-\sqrt{3}y)} \right. \nonumber\\
&   &\mbox{}\hspace{.20in}+ \left. D_{1\overline{1}0}e^{ik(3x/2-y\sqrt{3}/2)}+
D_{10\overline{1}}e^{ik(3x/2+y\sqrt{3}/2)}+
D_{01\overline{1}}e^{ik\sqrt{3}y} \right\} \nonumber \\
&   &\mbox{}+c.c.+h.o.t., \nonumber \end{eqnarray}
where $\epsilon \ll 1$ and $k$ is the wavenumber of the pattern.
The Fourier amplitudes are functions of the slow time $\tau_{2}=\epsilon^{2} t$.
The subscripts on the harmonics
$D_{lmn}$ indicate how the Fourier mode is expressed in terms of 
the
$O(\epsilon)$ modes; the first, second, and third subscripts indicate 
how many times $e^{ikx}$,
$e^{ik(-x/2+y\sqrt{3}/2)}$, and $e^{ik(-x/2-y\sqrt{3}/2)}$ are 
factors of the Fourier mode,
respectively. Bars over the subscripts indicate complex 
conjugates. For example, $D_{10\overline{1}}e^{ik(3x/2+y\sqrt{3}/2)}$
contains one $e^{ikx}$-mode and one complex conjugate of the 
$e^{ik(-x/2-y\sqrt{3}/2)}$-mode.

Inserting the expansion (\ref{eq:psihex}) into the model
equations (\ref{eq:mshe}) and (\ref{eq:mshe2}) and
eliminating the
terms involving $D_{lmn}$ at $O(\epsilon^{2})$,in each case one obtains at 
$O(\epsilon^{3})$ 
a set of amplitude equations of the form (\ref{eq:ampeq}) 
for the amplitudes $A$, $B$, and $C$.
In the case of the minimal model (\ref{eq:mshe}), the coefficients
of (\ref{eq:ampeq}) have the  values
\bea
f_{1}=3,\qquad
f_{2}=6, \qquad f_{3}=6,  \label{eq:coeffs}
\eea
and $R_{2} = (R-(k^2-1)^2)/\epsilon^{2}$.
Note that $R$ is expanded about the neutral curve for the specified hexagon wavenumber 
$k$, and not about $R=0$. To obtain a homogeneous scaling,
we have assumed the bifurcation to be weakly transcritical, and hence
\be 
\zeta = \alpha / \epsilon \equiv O(1). \label{eq:zeta}
\ee 
Note that the quadratic rotation term 
$\gamma\hat{e}_{k} \cdot (\nabla \psi \times \nabla 
(\nabla^{2} \psi))$
does not enter the amplitude equations (\ref{eq:ampeq})
if $\alpha\ll1$.
In terms of the original variables, the amplitude for the steady 
hexagon
solution is to this order given by 
\be A=B=C={\frac {-\alpha- \sqrt { 
\alpha^{2}             
-(R-(k^{2}-1)^{2})\,(f_{1}+f_{2}+f_{3})}}{(f_{1}+f_{2}+f_{3})}}.\label{eq:hexsoln}
\ee
In the case of the extended model (\ref{eq:mshe2}), the coefficients
of (\ref{eq:ampeq}) have the values 
\bea
\zeta&=&\left\{\alpha + \frac{\beta k^{2}}{2}\right\}/ \epsilon \equiv O(1), 
\label{eq:zeta2}\\
f_{1}&=&3 - \left\{\frac{9}{2(6k^{2}-15k^{4})}+
\frac{1}{k^{4}-2k^{2}}\right\}\beta^{2}k^{4},\label{eq:coeffs3}\\
f_{2,3}&=&6\mp g_{2}k^{4}\sqrt{3}
\mp \frac{2k^{6}\sqrt{3}}{4k^{2}-8k^{4}}\gamma\beta-
\left\{ \frac{4}{4k^{2}-8k^{4}}+\frac{1}{k^{4}-2k^{2}}\right\} \beta^{2}k^{4},
\label{eq:coeffs2}
\eea
The effect of
rotation is contained solely in $f_{2}$ and $f_{3}$ such that
$f_{2}(\gamma,g_{2})=f_{3}(-\gamma,-g_{2})$. Note that in 
(\ref{eq:coeffs3},\ref{eq:coeffs2}) $\alpha$ has been replaced by 
$\beta$ using (\ref{eq:zeta2}).  The hexagon solution is to this order given by 
\be A=B=C={\frac {-(\alpha+ {\frac{\beta k^{2}}{2}})- \sqrt { 
(\alpha+ {\frac{\beta k^{2}}{2}} )^{2}             
-(R-(k^{2}-1)^{2})\,(f_{1}+f_{2}+f_{3})}}{(f_{1}+f_{2}+f_{3})}}.\label{eq:hexsoln2}\ee
 
In either case, to study the stability of the hexagon solution $H(x,y)$ a general
Floquet ansatz with modulation wavevector $(\tilde{q},\tilde{p})$ is made,
\bea
\psi = H(x,y) + \delta e^{i\tilde{q}x+i\tilde{p}y} P(x,y,t),
\eea
where $P(x,y)$ has the periodicity of the hexagonal lattice and $\delta \ll 1$.
Being interested in the weakly nonlinear regime, we expand
$H(x,y)$ and $P(x,y)$ in $\epsilon$ and obtain
\begin{eqnarray} 
\psi & = &\epsilon 
\left\{ Ae^{ikx}+Be^{ik(-x/2+y\sqrt{3}/2)}+Ce^{ik(-x/2-y\sqrt{3}/2)} \right\}
\label{eq:psifinite}\\ 
                                   &   &\mbox{}+\epsilon \delta 
\left\{ P_{000}e^{i(qx+py)} \right\} \nonumber\\
                                   &   &\mbox{}+\epsilon^{2} 
\left\{ D_{000}+D_{200}e^{i2kx}+D_{020}e^{ik(-x+\sqrt{3}y)}+D_{002}e^{ik(-x-\sqrt{3}y)} 
\right. \nonumber\\                                   
&   
&\mbox{}\hspace{.20in} \left. +D_{1\overline{1}0}e^{ik(3x/2-y\sqrt{3}/2)}+
D_{10\overline{1}}e^{ik(3x/2+y\sqrt{3}/2)}+D_{01\overline{1}}e^{ik\sqrt{3}y} \right\} 
\nonumber\\                                     
&   &\mbox{}+\epsilon^{2} \delta 
\left\{ H_{0\overline{1}0}e^{i((q+k/2)x+(p-k\sqrt{3}/2)y)}+H_{100}e^{i((q+k)x+py)}+
H_{00\overline{1}}e^{i((q+k/2)x+(p+k\sqrt{3}/2)y)} \right. \nonumber\\  
&   &\mbox{}\hspace{.20in} \left. +H_{\overline{1}00}e^{i((q-k)x+py)}+
P_{010}e^{i((q-k/2)x+(p+k\sqrt{3}/2)y)}+P_{001}e^{i((q-k/2)x+(p-k\sqrt{3}/2)y)} \right\} 
\nonumber\\                                   
&   &\mbox{}+c.c.+h.o.t. \nonumber
\end{eqnarray}
with the amplitudes of the perturbation modes depending on time.
 Here the notation for the $O(\epsilon)$ and $O(\epsilon^{2})$ modes
is as before. 
The subscripts on the $O(\epsilon^{2}\delta)$ amplitudes indicate how 
many times 
the $e^{i(qx+py)}$-mode was multiplied by the
indicated Fourier components that make up the original hexagon 
solution to create the harmonic mode.
The reason for labelling some of the
$O(\epsilon^{2}\delta)$ modes as $H$ and some as $P$ will become apparent below.
In the expansion (\ref{eq:psifinite}) we have combined the modulation wavevector 
$(\tilde{q},\tilde{p})$ with the basic wavevector $(k,0)$
of the hexagon pattern into the perturbation wavevector $(q=k+\tilde{q},p=\tilde{p})$.
Thus, long-wave perturbations are characterized by $(q,p)$ being close to $(k,0)$.

As long as the wavevector $(q,p)$ of the perturbation mode $P_{000}$ is not
close to the wavevector of one of the hexagon modes the
harmonics $H_{lmn}$ and $P_{lmn}$ of $P_{000}$ are bounded away from the critical 
circle (cf. Fig.\ref{fig:ModeMap1} below). Thus, their damping is $O(1)$ and
they can be eliminated in favor of $P_{000}$ and the hexagon modes $A$, $B$, and
$C$. Consequently, they are small (of $O(\delta \epsilon^2)$) compared to $P_{000}$. 
This results in an evolution equation for $P_{000}$ alone.

It turns out that the expansion (\ref{eq:psifinite}) yields unsatisfactory results
for the growth rate of the perturbation as a function of 
$\theta$.
This is due to singularities that arise for $\theta=0$ and $\theta=\pm 60^{\circ}$.
For $k=\sqrt{q^2+p^2}=1$, as $\theta \rightarrow 0$ the perturbation modes
$P_{010}$ and $P_{001}$
approach the critical circle and their linear damping goes to zero. Thus, they 
cannot be eliminated adiabatically any more as signified by singularities in the
growth rate for $P_{000}$. Similarly, when $\theta \rightarrow 60^{\circ}$
the mode $H_{0\bar{1}0}$ approaches the critical circle.
The influence of the singularities at $0^{\circ}$ and $60^{\circ}$ on 
the eigenvalues  is so large that they become unreliable over the whole
range $0^{\circ} < \theta < 60^{\circ}$.

\newlength{\htw}
\setlength{\htw}{0.5\columnwidth}
\begin{figure*}[tbp]
 \centerline{\epsfxsize=0.8\htw{\epsfbox{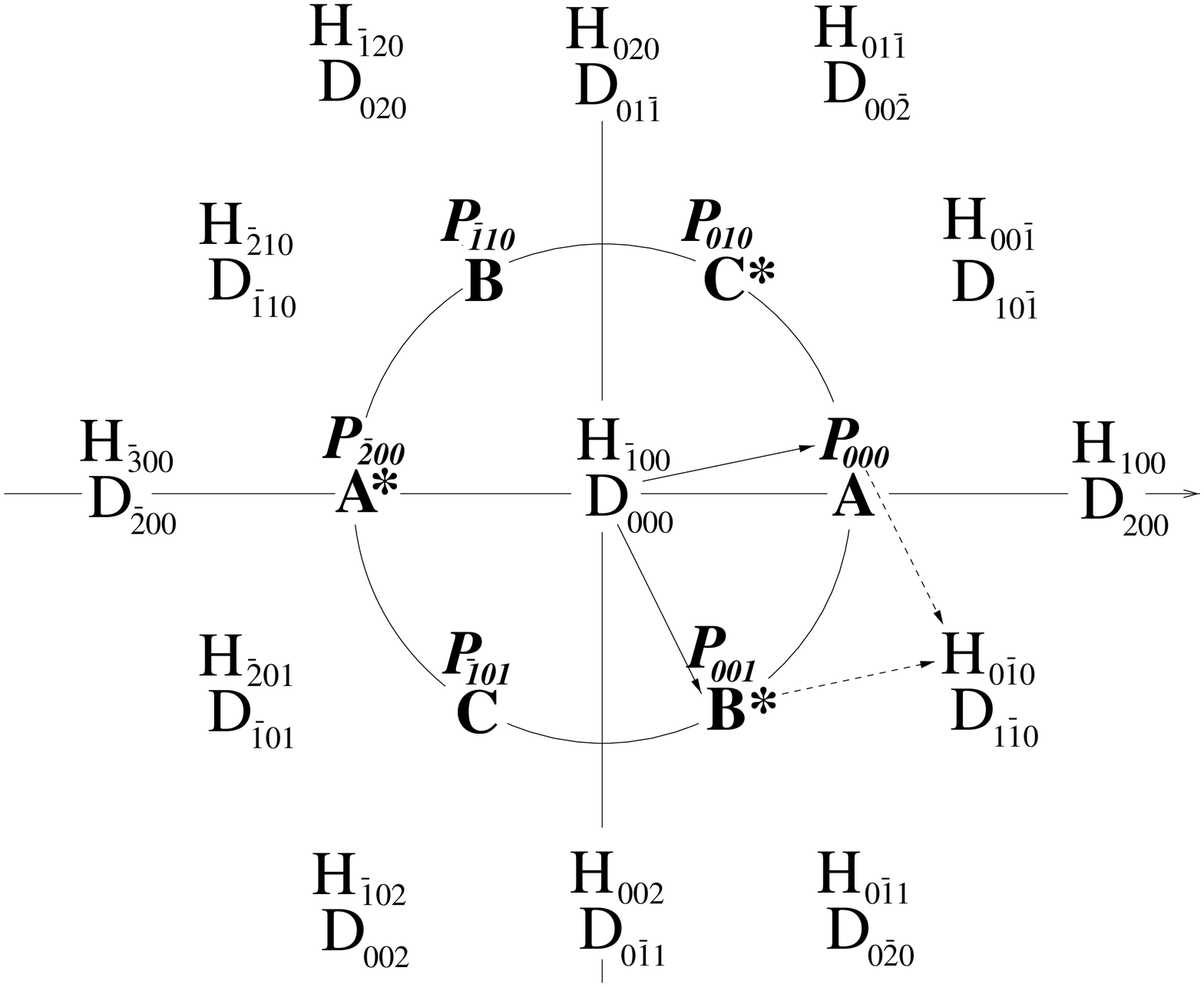}}
             \epsfxsize=0.8\htw{\epsfbox{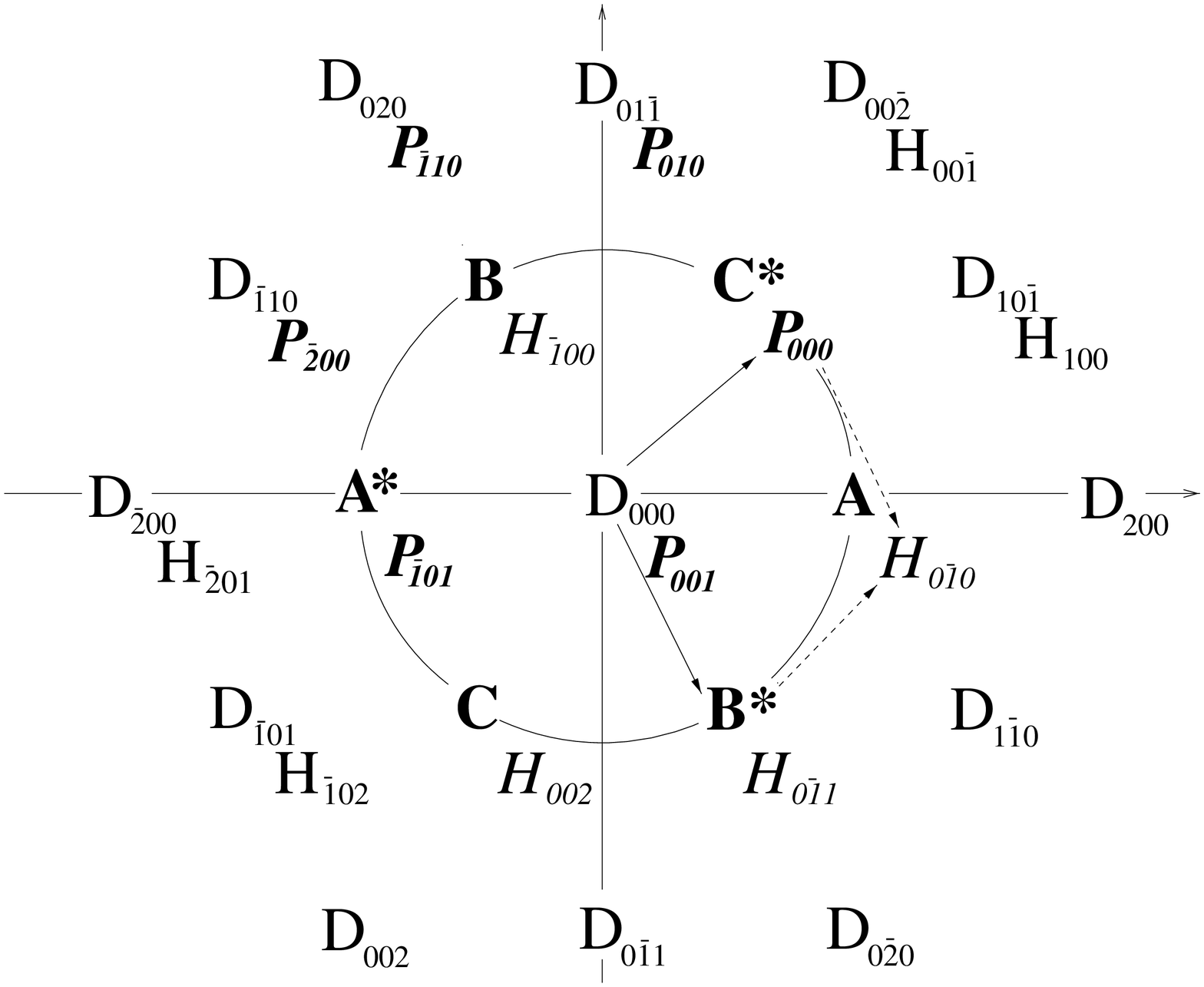}}
 }
\caption{Fourier space diagram showing the critical unit
circle and location of the modes used in the final expansion
of $\psi$ (cf. (\protect\ref{eq:psifinal})). 
Indicated with arrows is an example of the combination of an
$O(\epsilon\delta)$ perturbation mode and an $O(\epsilon)$
original hexagon mode driving a $O(\epsilon^{2}\delta)$ higher
harmonic mode.
\protect{\label{fig:ModeMap1}}}
\caption{Fourier space diagram where $\theta \rightarrow 60^{\circ}$.
Here the modes $H_{0\overline{1}0}$, $H_{0\overline{1}1}$, $H_{002}$,
and $H_{\overline{1}00}$ approach the critical unit circle despite
being of $O(\epsilon^{2}\delta)$, and thus drive a singularity
when solved for.
\protect{\label{fig:ModeMap2}}
}
\end{figure*}

In order to avoid the singularities, $P_{010}$ and $P_{001}$ are 
allowed to be of $O(\epsilon \delta)$,
i.e. of the same order as the original perturbation mode $P_{000}$.
Then at $O(\epsilon^{2} \delta)$ they drive the modes 
$P\,_{\overline{1}01}$ and $P\,_{\overline{1}10}$ leading 
to singularities in these modes
for $\theta \rightarrow 0$. Hence, modes $P_{\overline{1}01}$ and 
$P\,_{\overline{1}10}$ must also be allowed to be of $O(\epsilon \delta)$. 
In fact, all modes that come close to the  critical circle 
as $e^{i(qx+py)}$ approaches $e^{ix}$ must be kept at $O(\epsilon 
\delta)$.
They are labelled $P_{lmn}$.
This finally leads to the following expansion for $\psi$ (cf. Fig. \ref{fig:ModeMap1}),
\begin{eqnarray} \psi& = &\hspace{.20in}\epsilon 
\left\{ Ae^{ikx}+Be^{ik(-x/2+y\sqrt{3}/2)}+Ce^{ik(-x/2-y\sqrt{3}/2)} \right\}
\label{eq:psifinal}\\                     
&   &\mbox{}+\epsilon \delta \left\{ P_{000}e^{i(qx+py)}
+P_{010}e^{i((q-k/2)x+(p+k\sqrt{3}/2)y)}+P_{001}e^{i((q-k/2)x+(p-k\sqrt{3}/2)y)} \right. 
\nonumber\\                     
&   
&\mbox{}\hspace{.25in}+\left.P_{\overline{1}10}e^{i((q-3k/2)x+(p+k\sqrt{3}/2)y)}+
P_{\overline{1}01}e^{i((q-3k/2)x+(p-k\sqrt{3}/2)y)}+P_{\overline{2}00}e^{i((q-2k)x+py)} 
\right\} \nonumber\\
&   
&\mbox{}+\epsilon^{2} \left\{ D_{000}+D_{200}e^{i2kx}+D_{020}e^{ik(-x+\sqrt{3}y)}+
D_{002}e^{ik(-x-\sqrt{3}y)} \right. \nonumber\\                     
&   
&\mbox{}\hspace{.20in}\left.+D_{1\overline{1}0}e^{ik(3x/2-y\sqrt{3}/2)}+
D_{10\overline{1}}e^{ik(3x/2+y\sqrt{3}/2)}+D_{01\overline{1}}e^{ik\sqrt{3}y} \right\} 
\nonumber\\                        
&   &\mbox{}+\epsilon^{2} \delta 
\left\{ H_{\overline{1}00}e^{i((q-k)x+py)}+H_{100}e^{i((q+k)x+py)}+
H_{00\overline{1}}e^{i((q+k/2)x+(p+k\sqrt{3}/2)y)} \right. \nonumber\\
&   
&\mbox{}\hspace{.20in}+H_{0\overline{1}1}e^{i(qx+(p-k\sqrt{3})y)}+
H_{002}e^{i((q-k)x+(p-k\sqrt{3})y)}+H_{\overline{1}02}e^{i((q-2k)x+(p-k\sqrt{3})y)}
\nonumber\\                     
&   
&\mbox{}\hspace{.20in}+H_{\overline{2}01}e^{i((q-5k/2)x+(p-k\sqrt{3}/2)y)}+
H_{\overline{3}00}e^{i((q-3k)x+py)}+H_{\overline{2}10}e^{i((q-5k/2)x+(p+k\sqrt{3}/2)y)}
\nonumber\\                     
&   
&\mbox{}\hspace{.20in}+H_{\overline{1}20}e^{i((q-2k)x+(p+k\sqrt{3})y)}+
H_{020}e^{i((q-k)x+(p+k\sqrt{3})y)}+H_{01\overline{1}}e^{i(qx+(p+k\sqrt{3})y)}
\nonumber\\                     
&   
&\mbox{}\hspace{.20in}+\left.H_{0\overline{1}0}e^{i((q+k/2)x+(p-k\sqrt{3}/2)y)} 
\right\} +c.c.+h.o.t.  
\nonumber\end{eqnarray}

Certain values of $(q,p)$ drive a resonance that is still not 
accounted for in the scaling of the modes. For $k=1$, as $\theta$ approaches
$60^{\circ}$ the expansion breaks down again
because the modes corresponding to $H_{0\overline{1}0},
H_{0\overline{1}1},H_{\overline{1}00}$, and $H_{002}$ are now
near the critical circle
(cf. Fig. \ref{fig:ModeMap2}), but are still kept at $O(\epsilon^{2} \delta)$ in the
expansion (\ref{eq:psifinal}).
Similarly, the expansion breaks down as $\theta$ approaches 
$-60^{\circ}$. Due to the $60^{\circ}$-periodicity of the expansion, we may restrict
ourselves to $\theta$ between $-30^{\circ}$ and $30^{\circ}$,
and thus avoid having to do an altered calculation for these 
situations.
If $k=0.888$, 
 a resonance occurs already for $\theta=\pm 30^{\circ}$.
Since $k=.888$ is smaller than any hexagon wavenumber we 
consider, restricting $\theta$ to the range $-30^{\circ}$ to 
$30^{\circ}$ is adequate.

To summarize, in the expansion (\ref{eq:psifinal}) all modes are retained that are 
excited at $O(\epsilon^2)$ or $O(\epsilon^2\delta)$ by the perturbation modes or the
hexagon modes. This is sufficient to determine the growth rates of the perturbation 
modes at $O(\epsilon^3\delta)$. The expansion remains well ordered as long as the 
hexagon wavenumber and the perturbation wavenumber are close to 1 and the angle
$\theta$ is between $-30^\circ$ and $+30^\circ$. 

We now substitute (\ref{eq:psifinal}) into (\ref{eq:mshe}), expand, and 
use
equation (\ref{eq:hexsoln}) to find the perturbation growth rates for the minimal
model. For the extended model (\ref{eq:mshe2}) and (\ref{eq:hexsoln2}) are used.
To obtain a unified description
for the whole range of angles we
reconstitute the perturbation equations at $O(\epsilon^{2} \delta)$
and $O(\epsilon^{3} \delta)$ by substituting in for all scaled amplitudes, e.g.
$A={\cal A}/\epsilon$, $P_{lmn}={\cal P}_{lmn}/\epsilon \delta$, etc.,
as well as the scaled variables,  $\zeta=\alpha/\epsilon$ or $\zeta=(\alpha+\beta/2)/\epsilon$ 
(depending on the model) and
$R_{2}=(R-(k^{2}-1)^{2})/\epsilon^{2}$, and the scaled times,
$\partial_{T}=\partial_{t}+\epsilon\partial_{\tau_{1}}+\epsilon^{2}\partial_{\tau_{2}}$.
This leads to
\bea \partial_{T}{\cal P}_{000} & = &
(R-(q^{2}+p^{2}-1)^{2}){\cal P}_{000} \label{eq:stability} \\
 &   &\mbox{}+{\cal B}^{*}{\cal
P}_{010}f_{010}(q,p,k) +{\cal C}^{*}{\cal P}_{001}f_{001}(q,p,k) 
+\sum_{lmn} g_{lmn}(q,p,k) |{\cal A}|^{2} {\cal P}_{lmn}, \nonumber
\eea
where the sum goes over all six perturbation modes ${\cal P}_{lmn}$.
The equations for the other amplitudes ${\cal P}_{lmn}$ follow by cyclic permutation
within the circle of modes ($P_{000}$, $P_{010}$, $P_{\bar{1}10}$, $P_{\bar{2}00}$,
$P_{\bar{1}01}$, $P_{001}$).
It turns out that despite the assumption of weak transcriticality,
(\ref{eq:zeta}) or (\ref{eq:zeta2}),
the coefficients of the quadratic terms involving ${\cal B^*}$ and ${\cal C^*}$
 are not small in general.
While the coefficients go to zero for $\theta \rightarrow 0$ and
$\zeta=0$, they are of $O(1)$ for finite angles $\theta$.
For finite angles and $k$ close to $1$, however, the modes ${\cal P}_{010}$
and ${\cal P}_{001}$
to which ${\cal P}_{000}$ is coupled are small rendering the quadratic terms again
of the same order as the cubic terms.

%%%%%%%%%%%%%%%%%%%%%%%%%%%%%%%%%%%%%%%%%%%%%%%%%%%%%%%%%%%%%%%%%%%%%%%%%%%%%%%%%%%%%%%%%%%%%

\section{Linear Stability Results}
The general stability analysis (\ref{eq:stability}) leads to
$6$ eigenvalues. To provide a context we first discuss the special case
that the perturbation modes
coincide with the original hexagon modes (i.e. $(q,p) = (k,0)$).
The analysis is then equivalent to that of the amplitude equations (\ref{eq:ampeq}),
which can be rewritten as equations for three real
amplitudes and three phases. Due to translation symmetry in two directions, 
two of the phases decouple and
the eigenvalues corresponding to the translation modes are identically zero.
All of the relevant behavior is contained in the three
real amplitude equations and one phase equation. As long as the up-down symmetry is
broken sufficiently strongly the remaining phase
relaxes quickly to a stable fixed point,  and only three
real amplitude
equations are needed, corresponding to three eigenvalues instead of
six.
This system has been thoroughly studied by Swift and Soward \cite{Sw84,So85}.
In Fig.\ref{fig:bifdiags} the resulting bifurcation diagrams are sketched for the
case without and with rotation. 
The new feature introduced by the rotation is
a limit cycle of oscillating hexagons that branches off the hexagons and connects
it with the general solution
in which all three amplitudes have different magnitudes.
The limit cycle arises in a Hopf bifurcation at
\be
R_{Hopf} =  \frac{ 8 \epsilon^{2} \zeta^{2} 
(f_{2}+f_{3}+4f_{1}) }
                   { (f_{2}+f_{3}-2 f_{1})^{2} }
                   +(k^{2}-1)^{2},\label{eq:hopf}
\ee
with $\zeta$ given by (\ref{eq:zeta}) and (\ref{eq:zeta2}), respectively.
At $R_{Hopf}$ the real parts of two of the three amplitude  eigenvalues
become positive.
Without rotation the eigenvalues are real and the instability leads
through a transcritical, steady bifurcation directly to the unstable general solution
and renders the rolls the only stable state.
Note that because $f_{2}$ and $f_{3}$ only appear in the combination
$f_{2}+f_{3}$,
$R_{Hopf}$ depends on neither $\gamma$ nor $g_{2}$.
The same is true of the saddle-node bifurcation at which the hexagons come into
existence.
In fact, $\gamma$ and $g_{2}$ are relevant solely for the imaginary parts of
the eigenvalues.

\setlength{\htw}{0.5\columnwidth}
\begin{figure*}
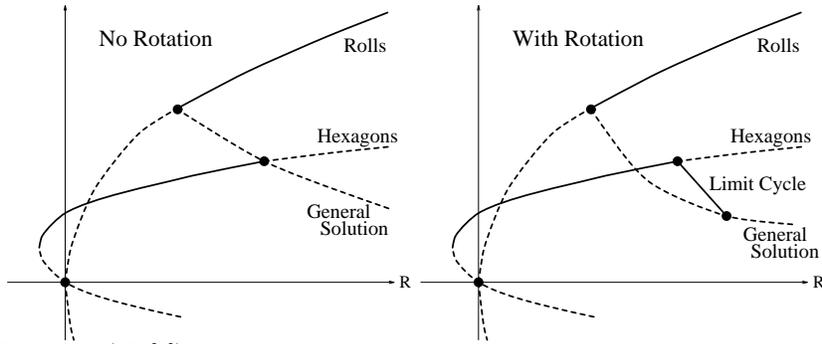

 \centerline{\epsfxsize=0.6\htw{\epsfbox{bif.no.rot.ps}}
             \epsfxsize=0.6\htw{\epsfbox{bif.with.rot.2.ps}}
  }
\caption{Bifurcation Diagrams
(cf. \protect\cite{Sw84}). The hexagon solution gains
stability through a saddle node bifurcation.
Without rotation the hexagons become unstable through a
transcritical bifurcation involving a general solution, where the
three amplitudes are all nonzero but not all equal.
With rotation the hexagons become
unstable to a limit cycle through a Hopf bifurcation. The
limit cycle in turn disappears through a collision with
the general solution.
\protect{\label{fig:bifdiags}}
}
\end{figure*}

In this paper we are interested in perturbation modes that differ in 
their wavevector from the
hexagon modes (i.e. $(q,p) \ne (k,0)$) and consider the dependence of the
eigenvalues on $(q,p)$. Thus, we deal with six branches of eigenvalues.
Away from the saddle-node bifurcation and for sufficiently strongly broken
up-down symmetry only four branches are relevant.
Two of them are associated with the instability
of hexagons to either the general solution (without rotation) or
the oscillating hexagons (with rotation).
We call the eigenvalues on these branches the `Hopf eigenvalues.'
The other two branches are associated with the translation modes
of the hexagons;
we call the eigenvalues on these branches `translation eigenvalues.'  Note
that they are identically zero only for $(q,p) = (k,0)$, 
i.e. for $\sqrt{q^2+p^2}=k$ and $\theta=0$. For other perturbation wavevectors they
are non-zero. In fact, 
in the presence of rotation the translation eigenvalues need not even
be real but can form a complex pair. In the following we discuss the dependence
of these four eigenvalues on the magnitude and, in particular,
on the orientation of the wavevector of the perturbation.
As discussed in the previous section, we restrict ourselves to
the range  $-30^{\circ}<\theta<30^{\circ}$
for the angle between the perturbation mode $P_{000}$ and mode $A$.

\setlength{\htw}{0.5\columnwidth}
\begin{figure*}[b]
 \centerline{\epsfxsize=0.8\htw{\epsfbox{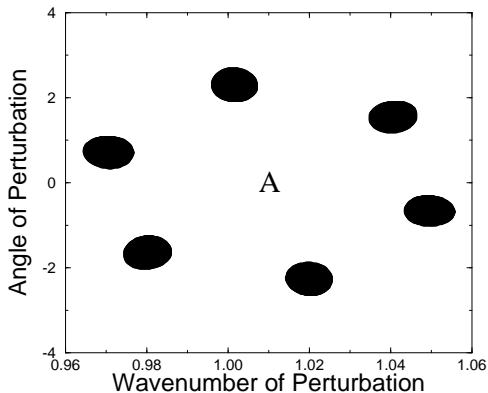}}}
\caption{Fourier space diagram with the shaded areas indicating
perturbations $P_{000}$ to which the hexagons in the minimal model are unstable for
$R=0.0294$, $k=1.01$, $\alpha=0.1$, $\gamma=6$.
The sixfold symmetry is due to the rotation symmetry. In this case the
instability first occurs at a finite angle $\theta$.
\protect{\label{fig:unstable}}
}
\end{figure*}

In Fourier space, the growth rates exhibit a sixfold symmetry
of areas of instability, centered around the wavevectors of the
hexagon modes $(k,\theta=2\pi n/6)$,
due to the rotation symmetry of the system (cf. \cite{SuTs94}).
Fig.\ref{fig:unstable} shows the instability regions for $P_{000}$ in the vicinity of
mode ${\cal A}$ for $k=1.01$.
In Fourier space (cf. Fig. \ref{fig:ModeMap1}), a rotation of ${\cal
P}_{000}$ about ${\cal A}$ by $60^{\circ}$ is 
equivalent to a rotation of the system about the origin by 
$60^{\circ}$ in the same direction.
The clockwise rotation of ${\cal P}_{000}$ about ${\cal A}$ by 
$60^{\circ}$ makes its
position equivalent to that of ${\cal P}_{010}$ before the rotation, 
the position
of ${\cal P}_{010}$ after the rotation equivalent to that of ${\cal 
P}_{\overline{1}10}$
before the rotation, and so on. 
Hence, the stability calculation exhibits a sixfold symmetry of 
unstable perturbation modes.
Additionally, for arbitrary perturbation wavenumber $(q,p)$,
the real part of the translation eigenvalues is not necessarily
zero when $\theta=0$; this is only true when
$\sqrt{q^{2}+p^{2}}=k$.

In the following we discuss the properties of the
minimal and the extended model separately. In both cases
we consider only those parameter values for which the
hexagons are stable to perturbations within the hexagonal lattice.
This means that we consider values of the bifurcation parameter
$R$ between the saddle-node and the Hopf bifurcation points.
Hence, the Hopf eigenvalues have negative real part for
$(q,p) = (k,0)$ in all our calculations.

\subsection{Stability Properties of the Minimal Model}

In the minimal model (\ref{eq:mshe}) we only found side-band 
instabilities involving the two branches of perturbations that are 
connected with the translation modes, i.e.  the growth rates of the 
perturbations vanish as the perturbation wavevector $(q,p)$ approaches 
that of a hexagon mode. 
For small $R$ the instabilities are long-wave 
and the wavevector of the most unstable perturbation is close to that of a 
hexagon mode, i.e. the instabilities arise at a small angle
$\theta$ between the perturbation modes and the hexagon modes.  The
eigenvalues are real for small rotation rates, but become complex as the
rotation rate is increased.

Further away from threshold the instabilities
become short-wave.  This is shown in Fig.\ref{fig:minimal.transinstab} where the real
part of the growth rate of the least stable modes is shown as a 
function of the angle $\theta$ of the perturbation modes.  In this 
illustration the magnitude of the perturbation wavevector is chosen to 
be equal to that of the hexagons, $\sqrt{q^{2}+p^{2}}=k$.  For small 
angles the perturbations are damped.  Only for larger angles the real 
part of the growth rate changes sign (dashed line).  In this case the 
relevant eigenvalues are complex.  It is worth mentioning that the 
translation mode and the Hopf mode (solid line) influence each other 
in this case for small angles.  This will become more important in the 
extended model (cf. Fig.\ref{fig:branchswitch} below).

\setlength{\htw}{0.5\columnwidth}
\begin{figure}[h]
 \centerline{\epsfxsize=0.8\htw{\epsfbox{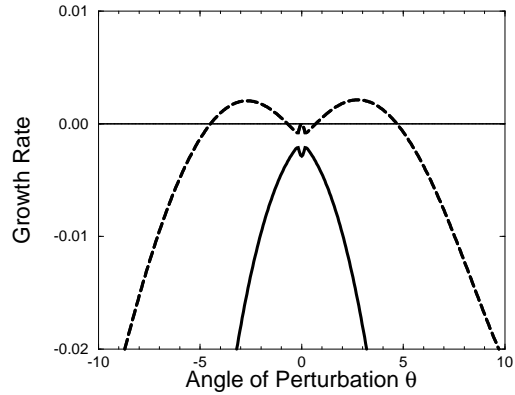}}}
\caption{A short-wave instability driven by the translation modes in 
the minimal model for $\alpha=0.1$, $\gamma=6$, $R=0.04$, and 
$k=1.01=\protect\sqrt{q^{2}+p^{2}}$.
}
\label{fig:minimal.transinstab}
\end{figure}

\setlength{\htw}{0.5\columnwidth}
\begin{figure}[h]
 \centerline{\epsfxsize=0.6\htw{\epsfbox{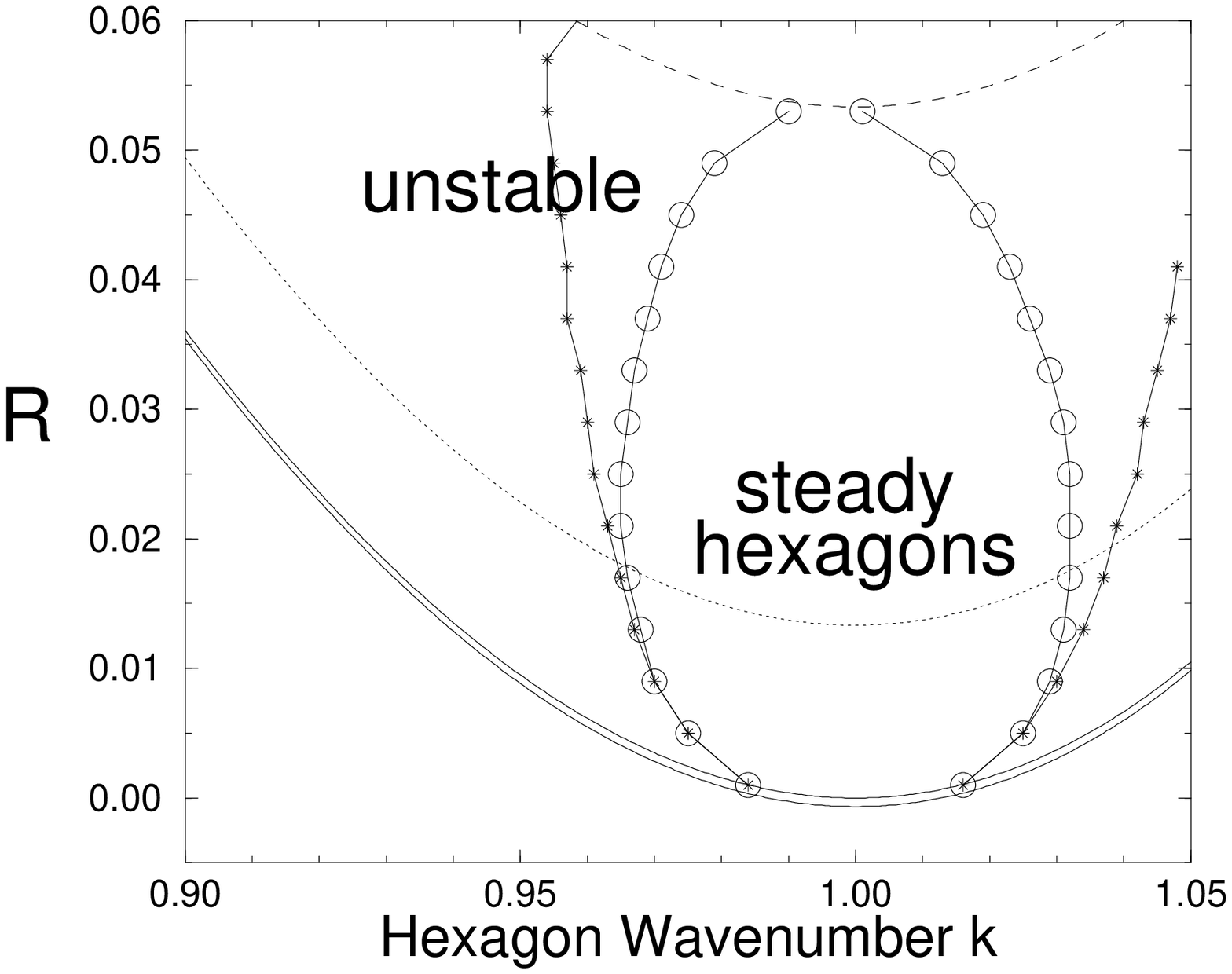}}
             \epsfxsize=0.6\htw{\epsfbox{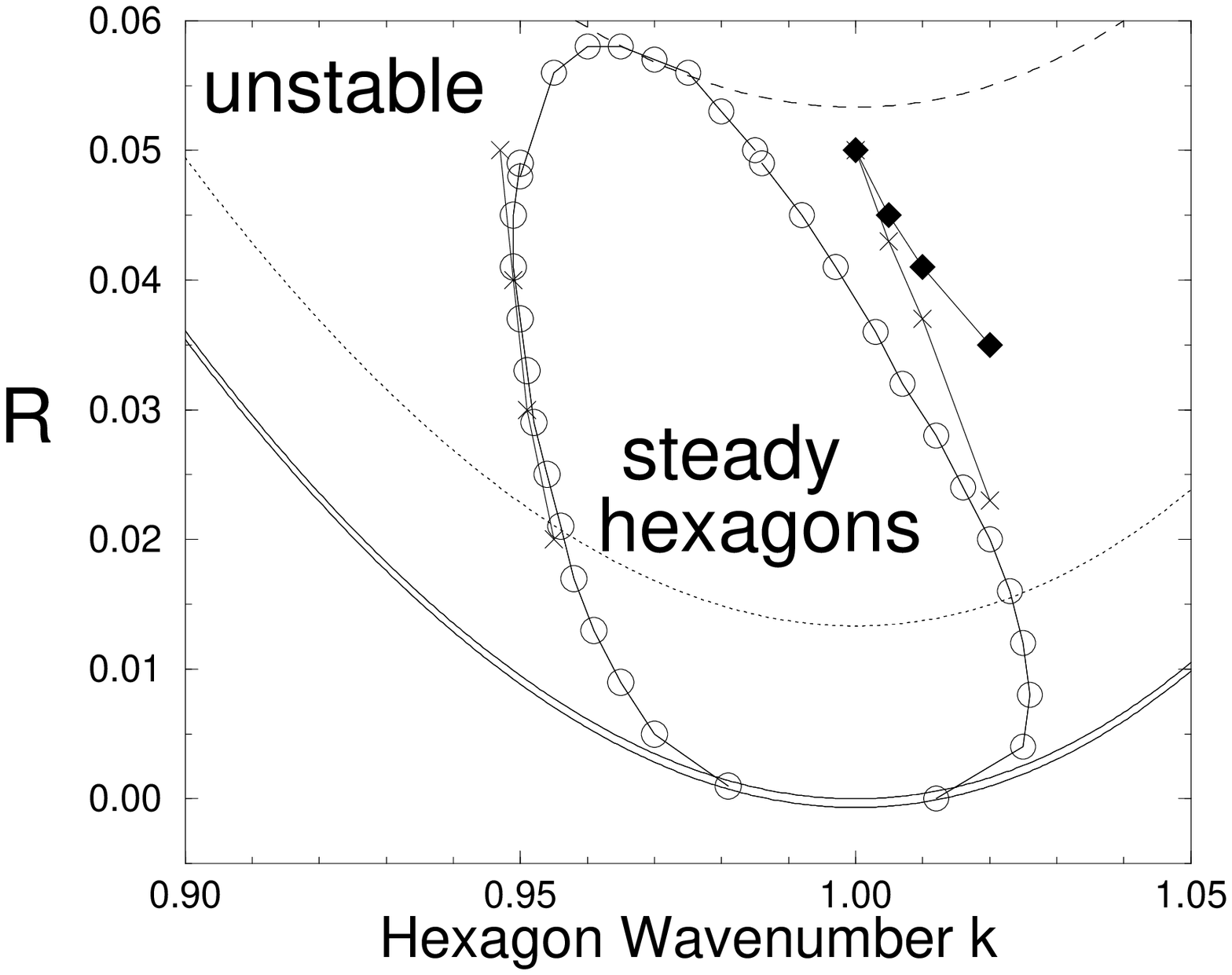}}
             \epsfxsize=0.6\htw{\epsfbox{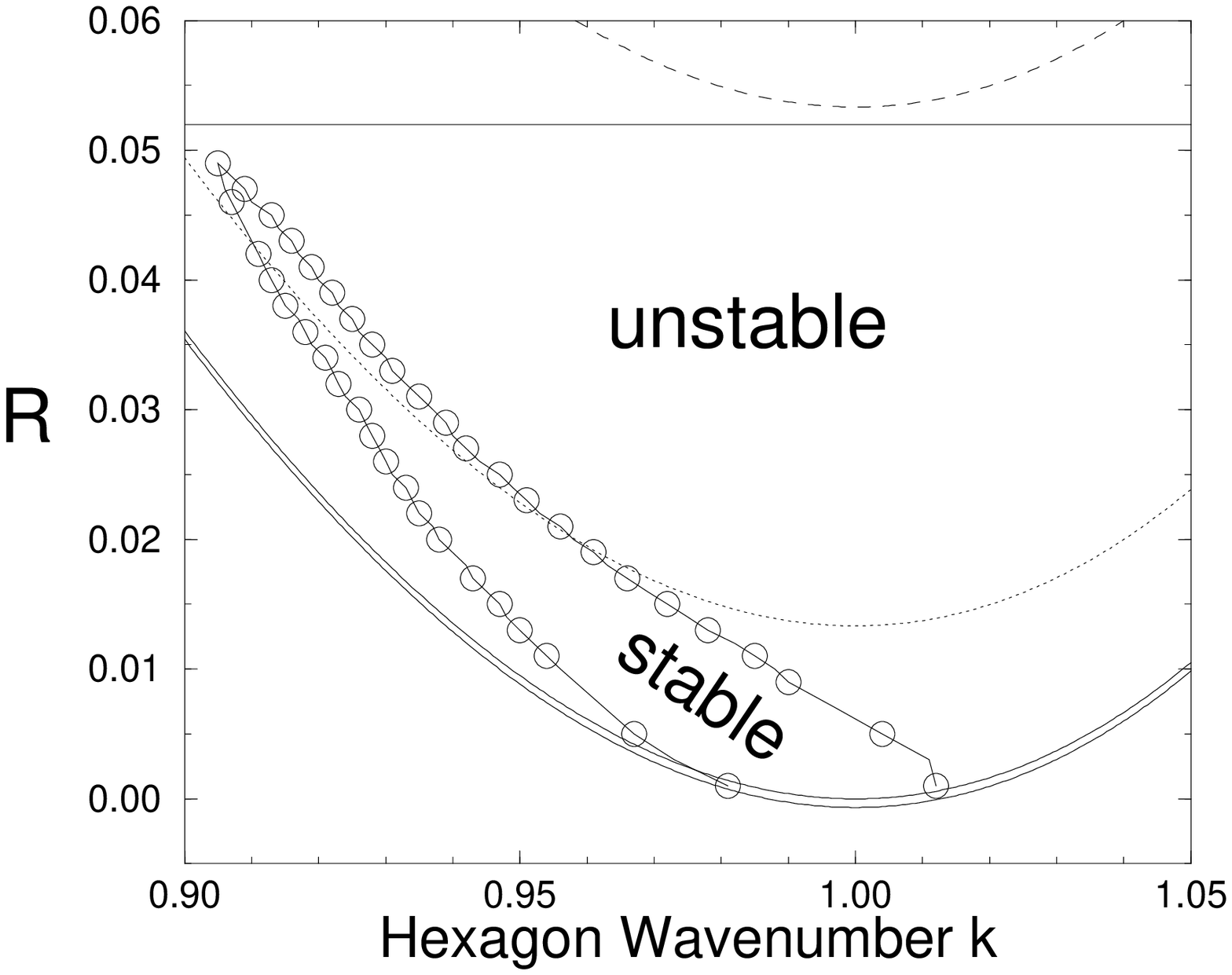}}}
\caption{
Stability regions   for the minimal model for various values of $\gamma$.
Open circles denote sideband instability in all cases.
Solid lines: neutral
curve and saddle-node bifurcation.
Dashed line: instability of hexagons to rolls.
Dotted line: lower stability boundary of the rolls (cf.
Fig. \protect\ref{fig:bifdiags}).
{\it a)} $\gamma=2$. Solid
symbols denote long-wave instability.
{\it b)} $\gamma=6$. Crosses
denote numerical instability of hexagons, diamonds denote stability limit
of modulated oscillations.
{\it c)} $\gamma=14$. Line
denotes value of $R$ of numerical investigation (see 
Fig.\protect\ref{fig:disorgap} below).
}
\label{fig:a.g.vary.gamma}
\end{figure}

The stability limits obtained from (\ref{eq:stability})
are shown in Fig.\ref{fig:a.g.vary.gamma} for a 
range of rotation rates $\gamma$. In these calculations the perturbation 
wavenumber  is allowed to vary within the range $0.9\le \sqrt{q^2+p^2}\le 1.1$ and the
angle within the range $-30^\circ \le \theta \le 30^\circ$. 
We note that variation of the non-rotational quadratic terms, in
this case $\alpha$, simply varies the strength of the subcriticality
of the hexagon solution and shifts the instability of the hexagons to rolls
up and down, thus setting the over-all scale for the stability range of the
hexagons; no qualitative changes in the stability limits were found.
The solid lines in Fig.\ref{fig:a.g.vary.gamma} indicate the neutral
curve and the locus where the hexagons first arise in a saddle-node
bifurcation.
The dashed line indicates where the hexagons become
unstable to rolls ($via$ the unstable, general solution in which all
three modes $A$, $B$, and $C$ have different amplitudes 
(Fig. \ref{fig:bifdiags})).
In the generic case in which the rotation affects the difference between the 
cubic coupling coefficients $f_{2}$ and $f_{3}$ this instability 
is oscillatory and leads to the
oscillating hexagons. The dotted line indicates the lower stability
limit of rolls with respect to hexagons. The side-band instability
of the hexagons is shown with open circles. Its long-wave limit is
given with small, solid symbols. As mentioned above, the long-wave 
result gives the correct stability limit for small $R$. For larger 
$R$, however, the short-wave instability sets in first. As the 
rotation rates is increased the stability limits become more 
asymmetrical as shown in Fig.\ref{fig:a.g.vary.gamma}b. 
Also shown in this case (as crosses) are the
stability limits as they are obtained in full numerical simulations 
of (\ref{eq:mshe}). For sufficiently large rotation
rates the stability region of the hexagons detaches from the line 
that indicates the transition to the general solution. Thus, over a range
of the control parameter $R$ the hexagons are still stable with 
respect to rolls, but nevertheless at no wavenumber are the hexagons 
stable with respect to side-band instabilities. This suggests the 
possibility of persistent dynamics, which is discussed below.

\subsection{Stability Properties of the Extended Model}

Within the extended model (\ref{eq:mshe2})  the coefficients $f_{2}$ and $f_{3}$
differ from each other for non-vanishing rotation rates and in the absence of
side-band perturbations hexagons become unstable to oscillating 
hexagons with increasing $R$ (dashed line in
Figs. \ref{fig:varyg2},\ref{fig:varyg2.only}b).
With side-band perturbations included it turns out that this Hopf
bifurcation can occur first for finite angles between the perturbation 
mode and the hexagon mode, as shown in Fig.\ref{fig:branchswitch}a. Thus, depending on
parameters the hexagons can become unstable with respect to modes 
rotated relative to the hexagons in two different ways: the 
destabilizing modes can be connected with the translation modes or 
the Hopf mode. There are two possibilities for the
transition between the two regimes to occur. Either the two modes interchange their roles
without affecting each other, or the transition involves a
switching between the two branches. The latter is illustrated in 
Fig.\ref{fig:branchswitch}. While for $R=0.035$ and $k=0.950$ the destabilizing mode is clearly
associated with the Hopf mode (Fig.\ref{fig:branchswitch}a),
it is clearly connected with the
translation branch for $R=0.020$ and $k=0.957$ (Fig.\ref{fig:branchswitch}c).
The transition occurs around
$R=0.028$ and $k=0.953$ for which value the Hopf and the translation-mode branch
merge and the destabilizing mode cannot uniquely be attributed to either branch
(Fig.\ref{fig:branchswitch}b).
For these parameter values the translation modes and the Hopf mode are in resonance,
i.e. they have the same frequency.
Fig.\ref{fig:varyg2} shows where the stability limits are due to the Hopf 
branch and to the translation-mode branch, respectively. 
Comparing Figs.\ref{fig:branchswitch}a and \ref{fig:branchswitch}c
one may expect that the two types of 
instabilities would lead to different behavior. This is, however, not 
the case (see sec.\ref{sec:numeric} below).

\vspace{.30in}
\setlength{\htw}{0.5\columnwidth}
\begin{figure*}[b]
 \centerline{\epsfxsize=0.6\htw{\epsfbox{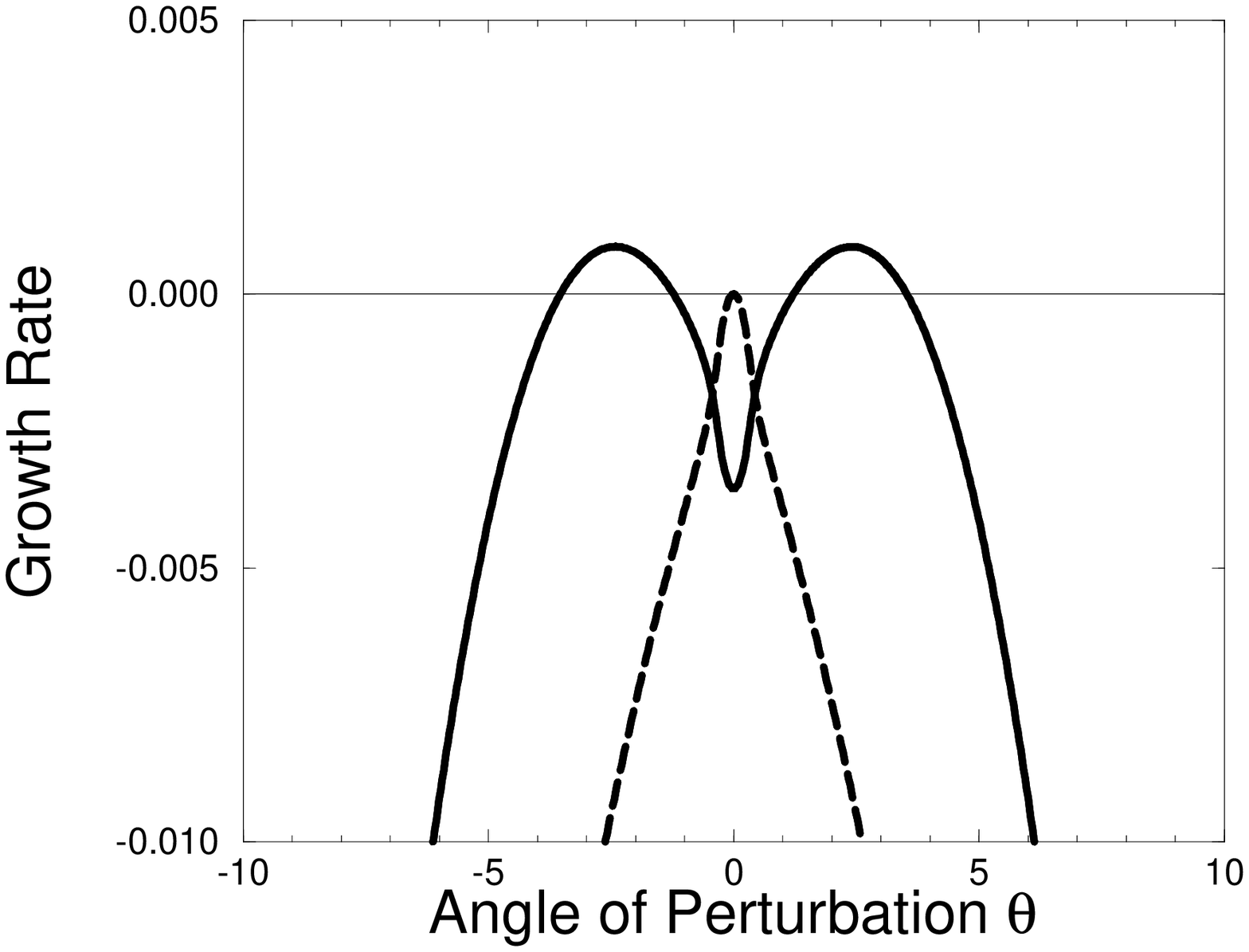}}
             \epsfxsize=0.6\htw{\epsfbox{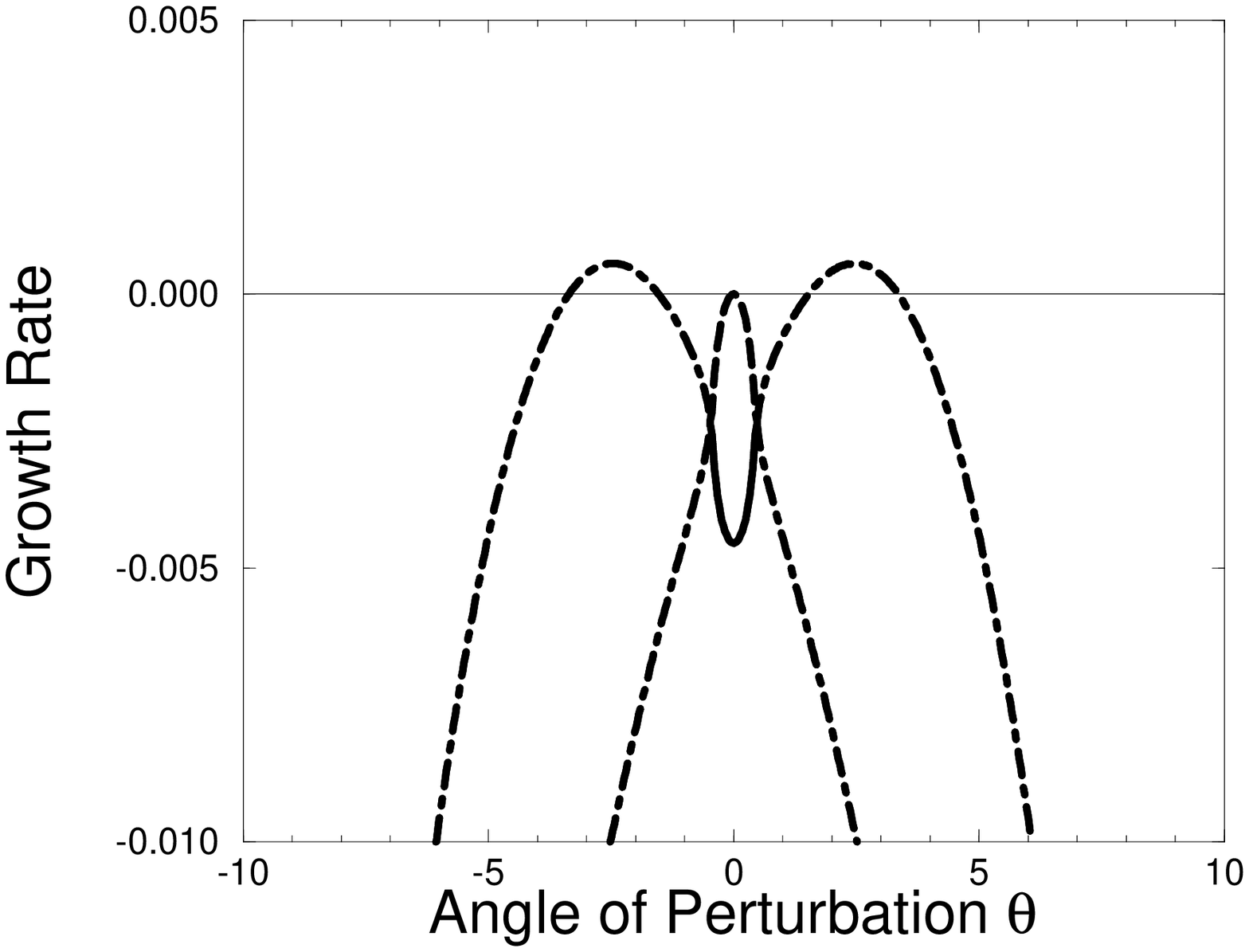}}
             \epsfxsize=0.6\htw{\epsfbox{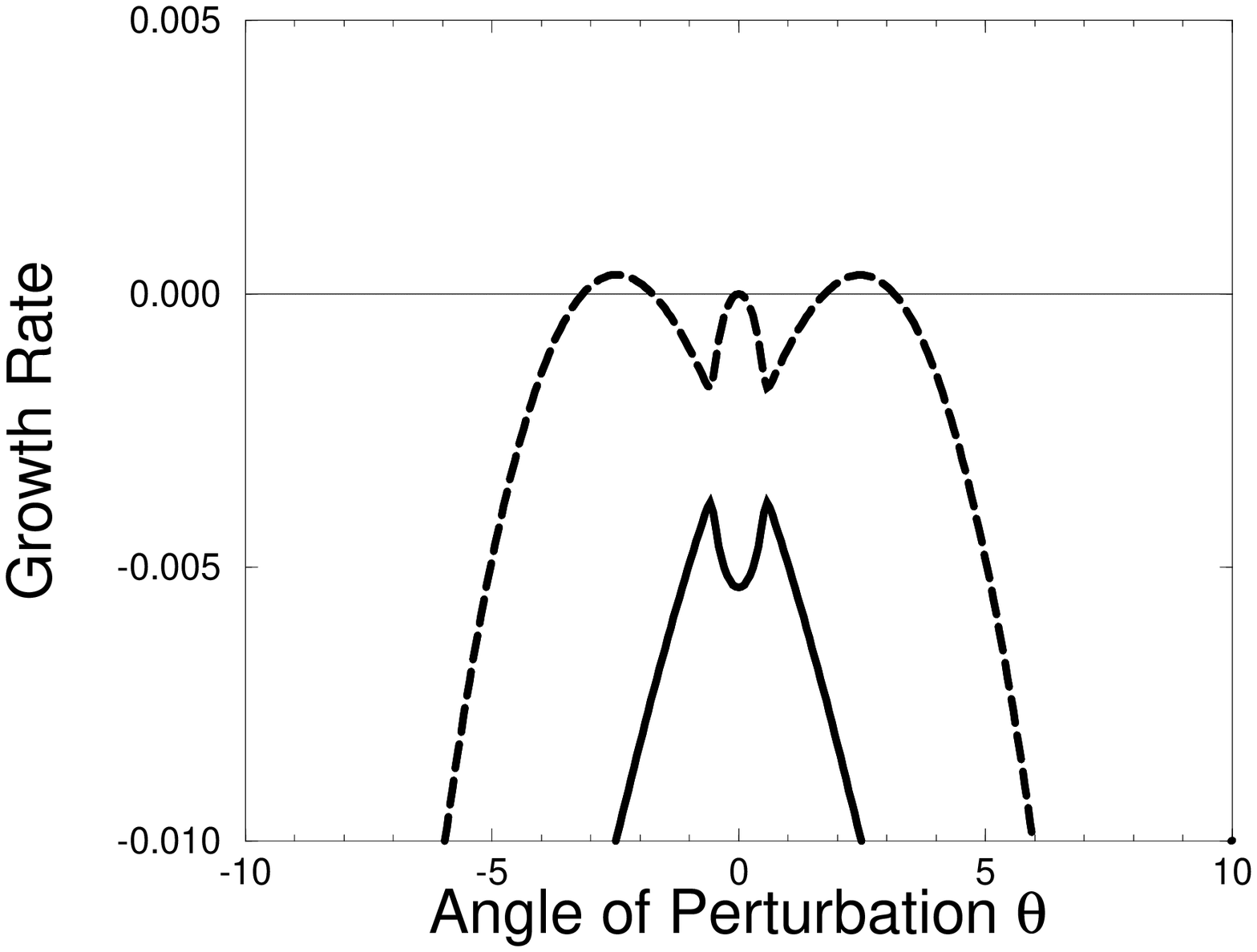}}
}
\caption{The branch switching transition in the extended model for
$\alpha=0$, $\beta=0.2$, $\gamma=6$, $g_{2}=1$.
{\it a)} $R=0.035$,
$k=0.950=\protect\sqrt{q^{2}+p^{2}}$.
An oscillatory instability driven by the Hopf modes.
{\it b)} $R=0.028$,
$k=0.953=\protect\sqrt{q^{2}+p^{2}}$.
The Hopf and translation
eigenvalues meet at two points; in this situation it is not clear
which dash-dotted eigenvalue branch
belongs to the Hopf and which to the translation eigenvalues.
{\it c)} $R=0.020$,
$k=0.957=\protect\sqrt{q^{2}+p^{2}}$.
An oscillatory instability driven by the translation  modes (cf. Fig.\protect\ref{fig:minimal.transinstab}).
\protect{\label{fig:branchswitch}}
}
\end{figure*}

In the absence of rotation, all eigenvalues are real.
As rotation is increased, the translation eigenvalues become complex 
for a range of $\theta$.
In  Fig.\ref{fig:varyg2.only}a a case is shown in which the translation eigenvalues
are a complex conjugate pair with negative real part for small $|\theta|$, but
for larger $|\theta|$ the eigenvalues split and turn real
with one eigenvalue becoming positive driving a steady instability.
Here the only rotation term is $g_{2}$. Thus, for $g_2=0$ there is no rotation
in the system and all eigenvalues are real. One can get qualitatively 
similar instabilities with only $\gamma$ providing the rotation. However, 
rotation driven by $\gamma$ does not generally have a real eigenvalue becoming 
positive for such a large $|\theta|$ as shown in Fig.\ref{fig:varyg2.only}a.
Fig.\ref{fig:varyg2.only}b shows the resulting stability limits. As in the cases
discussed above, close to threshold 
the limits are given by the long-wave perturbations (small filled 
symbols), while further 
above threshold the relevant instabilities become short-wave. In 
contrast to the cases discussed before, however, the short-wave 
instability now involves a single real eigenvalue.

\setlength{\htw}{0.5\columnwidth}
\begin{figure}[h]
 \centerline{
	     \epsfxsize=0.6\htw{\epsfbox{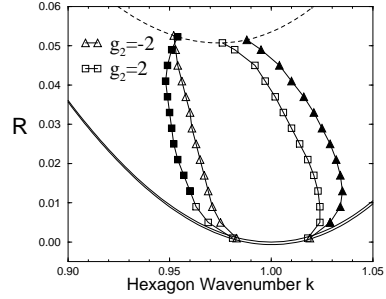}}
	     }
\caption{Stability regions 
for various values of $g_{2}$ with $\alpha=0$, 
$\beta=0.2$, $\gamma=6$.
Solid lines: neutral stability and hexagon saddle-node bifurcation curves.
Dashed line: Hopf bifurcation to oscillating hexagons.
Solid symbols: instability associated with the Hopf mode,
open symbols: instability associated with the translation mode.
\protect{\label{fig:varyg2}}
}
\end{figure}

\setlength{\htw}{0.5\columnwidth}
\begin{figure}[h]
 \centerline{
             \epsfxsize=0.6\htw{\epsfbox{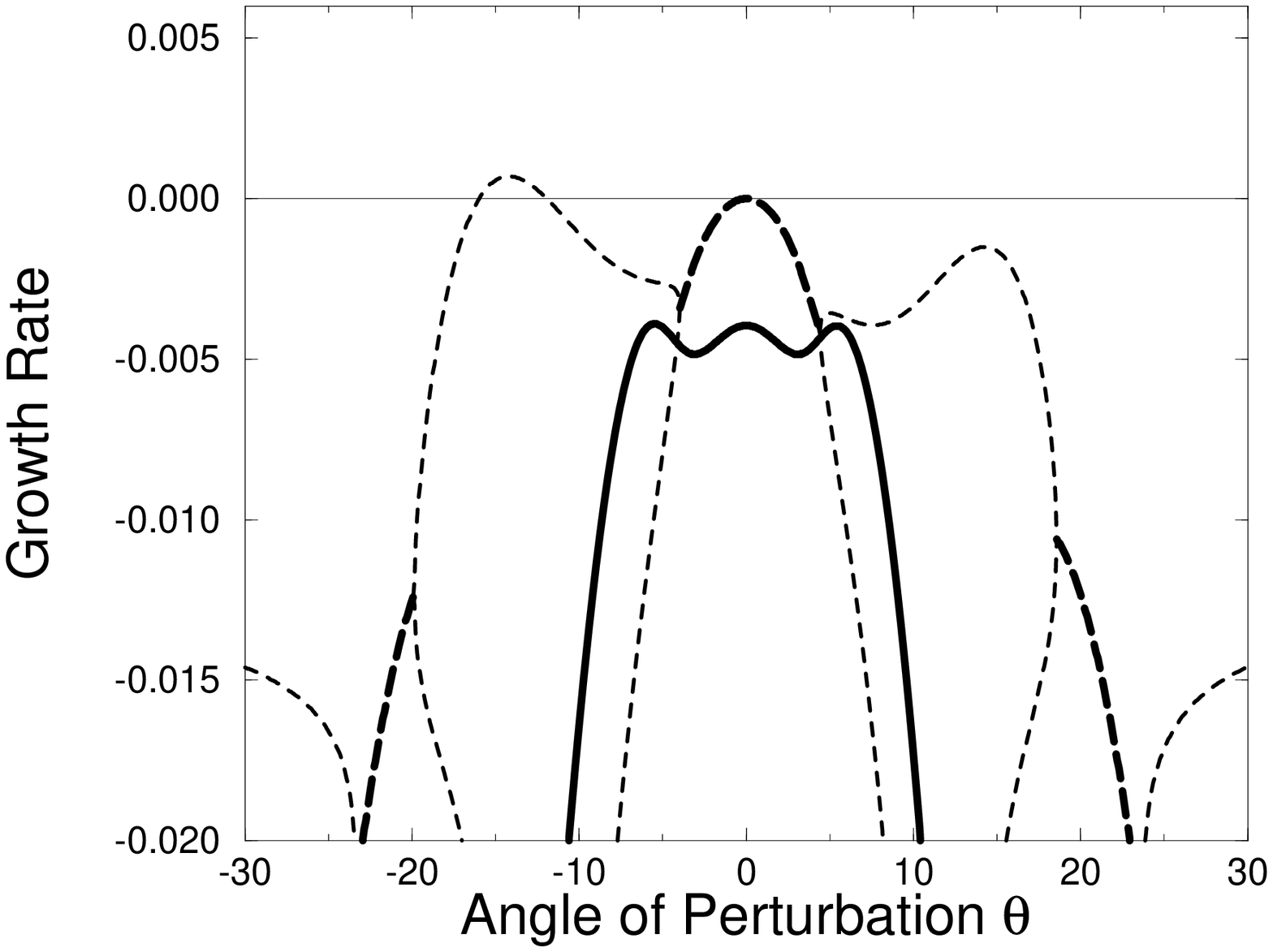}}
             \epsfxsize=0.6\htw{\epsfbox{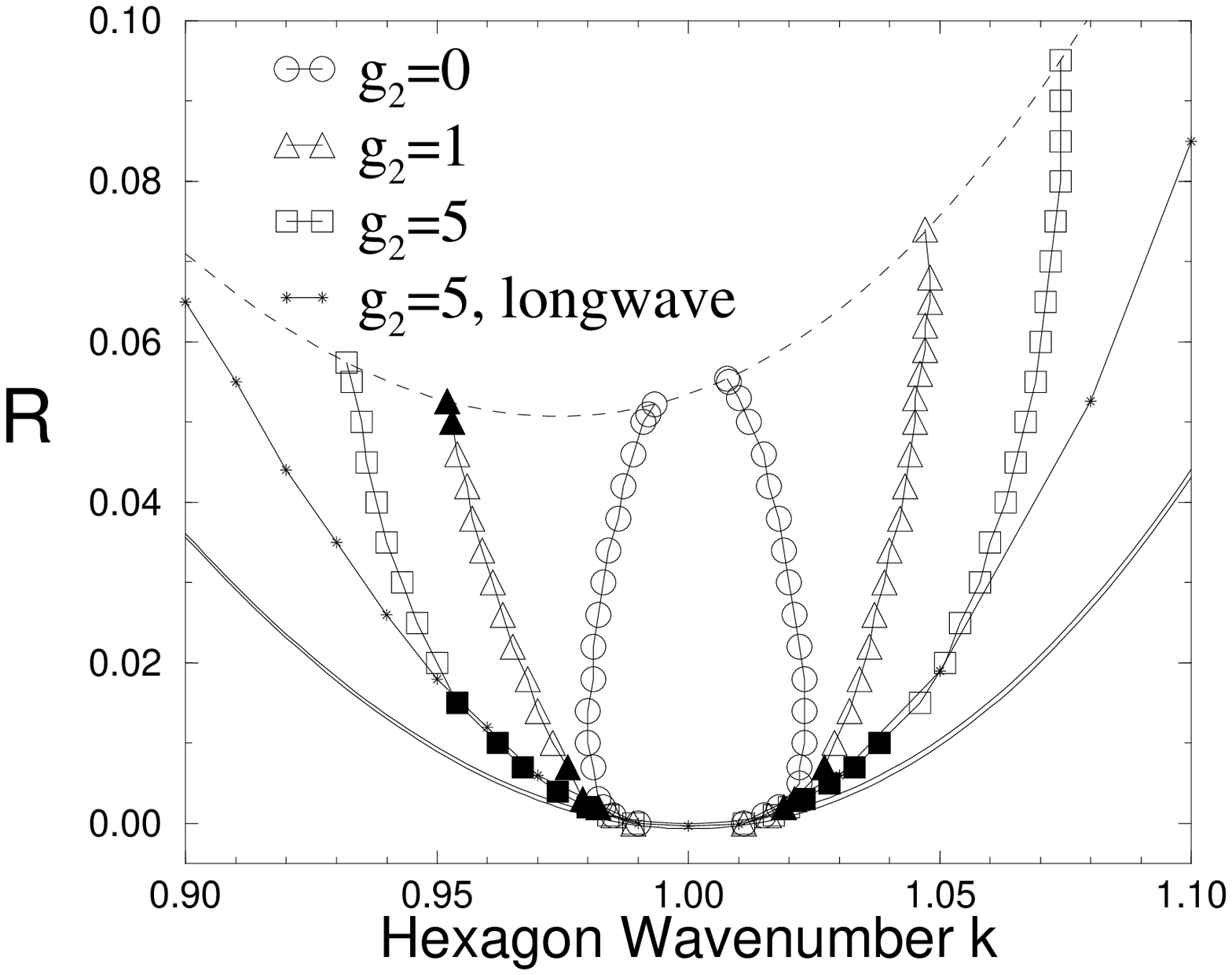}}
	     }
\caption{{\it a)} One of the real eigenvalues (thin dashed line) resulting
from the splitting of the complex conjugate pair (thick dashed line)
of translation eigenvalues is positive for
$R=0.035$, $k=0.938=\protect\sqrt{q^{2}+p^{2}}$, $\alpha=0$,
$\beta=0.2$,
$\gamma=0$, $g_{2}=5$.
Note that individual eigenvalues can change from being real to being
part of a complex conjugate pair of eigenvalues as $\theta$
is varied.
{\it b)} Stability regions
for various values of $g_2$ with $\alpha=0$, $\beta=0.2$, $\gamma=0$. 
Solid symbols: oscillatory instability; open symbols instability 
involving a real eigenvalue. 
\protect{\label{fig:varyg2.only}}
}
\end{figure}
 
%%%%%%%%%%%%%%%%%%%%%%%%%%%%%%%%%%%%%%%%%%%%%%%%%%%%%%%%%%%%%%%%%%%%%%%%%%
\section{Numerical Simulations}

\label{sec:numeric}
To support the analytical calculation, and to study the nonlinear
behavior
ensuing from the instabilities,  numerical simulations of
(\ref{eq:mshe}) and (\ref{eq:mshe2})
were performed.
A Runge-Kutta method with an integrating factor that computes
the linear derivative terms exactly was used.
Derivatives were computed in Fourier space, using a two-dimensional
complex
fast Fourier transform (FFT).

The numerical simulations were done in a rectangular $32\pi$ by
$64\pi/\sqrt{3}$
box of aspect ratio $2/\sqrt{3}$ with periodic boundary conditions.
The aspect ratio of $2/\sqrt{3}$ was used to allow for regular
hexagonal patterns.
For the small amplitudes investigated here it was sufficient to use a
128 by 128
lattice keeping only up to the third harmonics of the modes
on the critical circle.
This allowed a larger number of modes within the critical annulus,
reducing the effects of anisotropy.

In all cases investigated, if the instability is associated with a real eigenvalue
it does not saturate and eventually leads again to steady hexagons with a 
wavevector in the
stable region. This is true for the usual long-wave case close to onset as well as for the
case when the instability arises first at a finite modulation wavevector
(cf. Fig.\ref{fig:varyg2.only}a).

The oscillatory instabilities exhibit more interesting behavior.
While they also can take the hexagon pattern back into the stable wavenumber
band, over a range of parameters
they lead in a supercritical bifurcation to
modulated hexagons. They 
are quite similar to the oscillating hexagons identified earlier
by Swift \cite{Sw84} and Soward \cite{So85} in that the hexagons are temporally 
modulated in time with modes corresponding to different orientations being phase-shifted
with respect to each other. In the modulated hexagons the oscillations occur,
however, predominantly in three side-band perturbation modes, which implies also a
spatial modulation of the hexagon pattern. 
A typical time sequence over one period $T=225$ 
is displayed in Fig.\ref{fig:oschex}.
Since the wavevector $(q,p)$ of the sideband perturbation
is close to that of the steady hexagons only one modulation wavelength fits into the
system. Fig.\ref{fig:oschex}
shows that the spatio-temporal modulation is in the form of a standing rather than
a traveling wave.
The modulated hexagons are stable in a small region beyond
the stability limit of the steady hexagons as delimited by the crosses and diamonds
in Fig.\ref{fig:a.g.vary.gamma}b and in Fig.\ref{fig:detail}. 

\setlength{\htw}{0.5\columnwidth}
\begin{figure}[h]
 \centerline{
             \epsfxsize=0.6\htw{\epsfbox{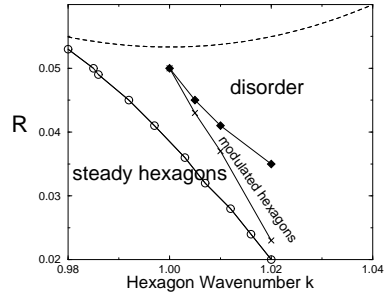}}
  }
\caption{A detail of stability diagram Fig.\protect\ref{fig:a.g.vary.gamma}b.
Circles: stability boundary of hexagons from weakly nonlinear analysis;
crosses: numerical stability boundary of hexagons on the 128 by 128 lattice;
solid diamonds: stability boundary of the modulated hexagon state.
}
\label{fig:detail}
\end{figure}
\setlength{\htw}{0.5\columnwidth}
\begin{figure}[h]
 \centerline{\epsfxsize=0.5\htw{\epsfbox{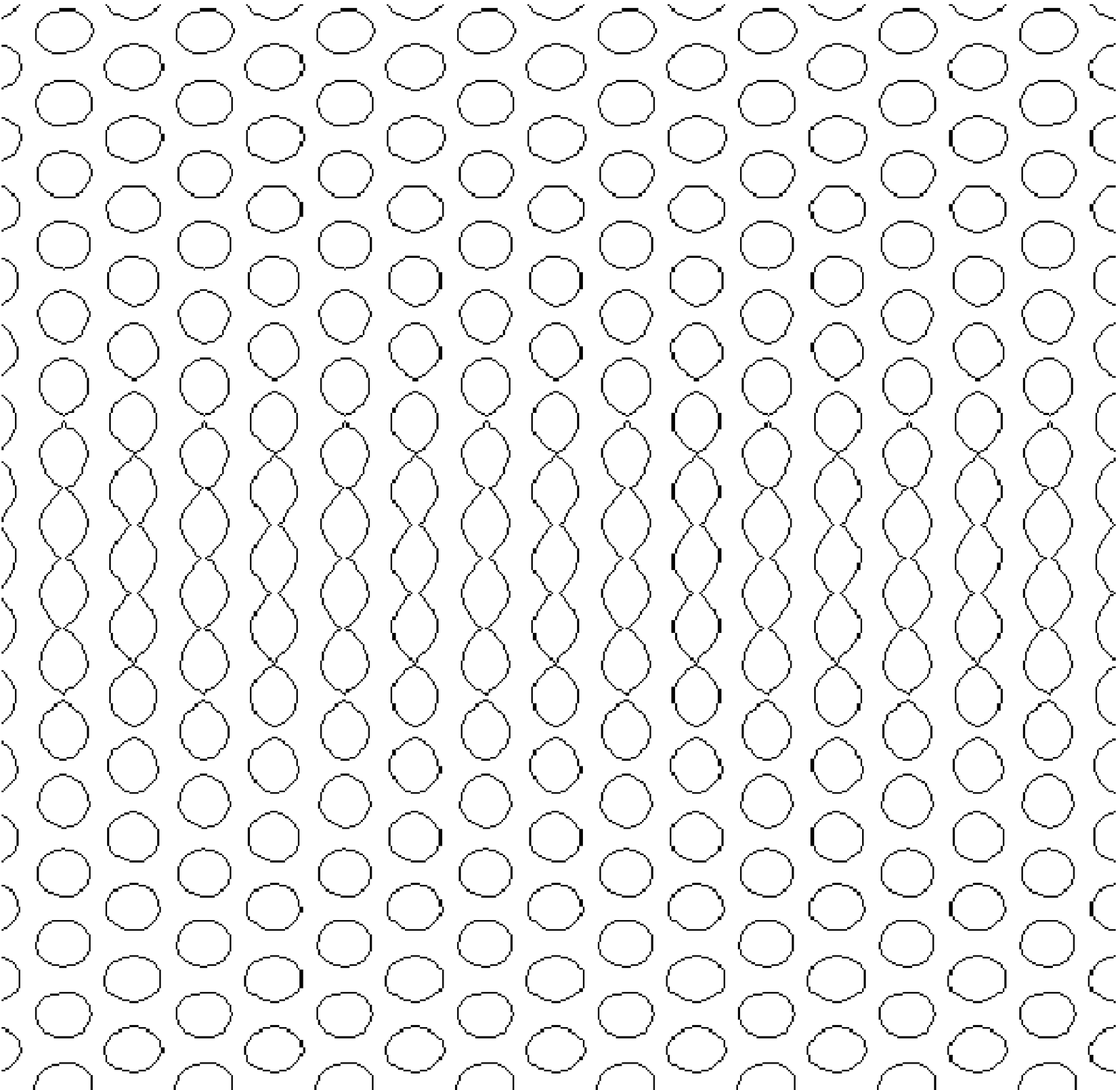}}
             \epsfxsize=0.5\htw{\epsfbox{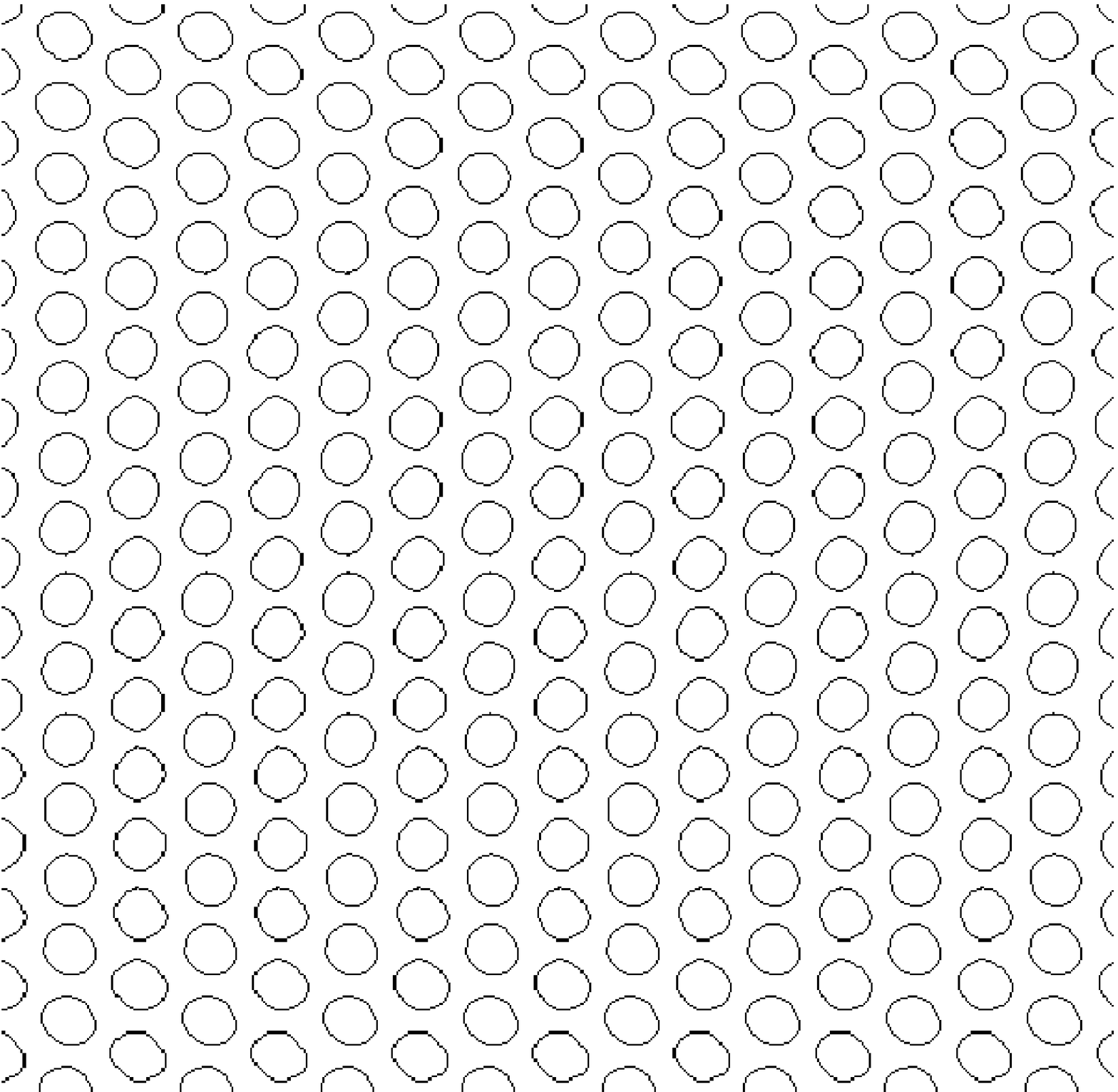}}
             \epsfxsize=0.5\htw{\epsfbox{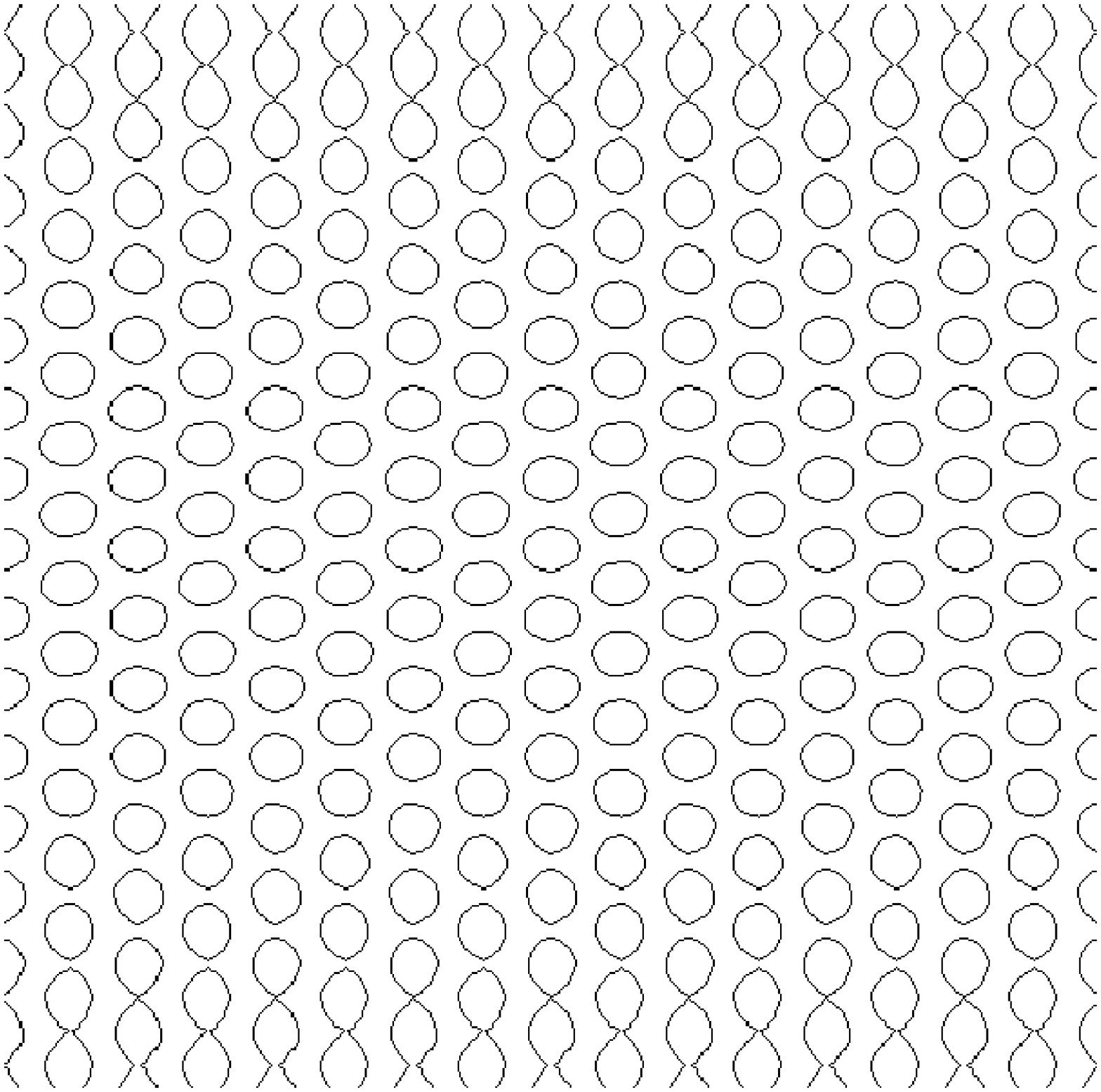}}
             \epsfxsize=0.5\htw{\epsfbox{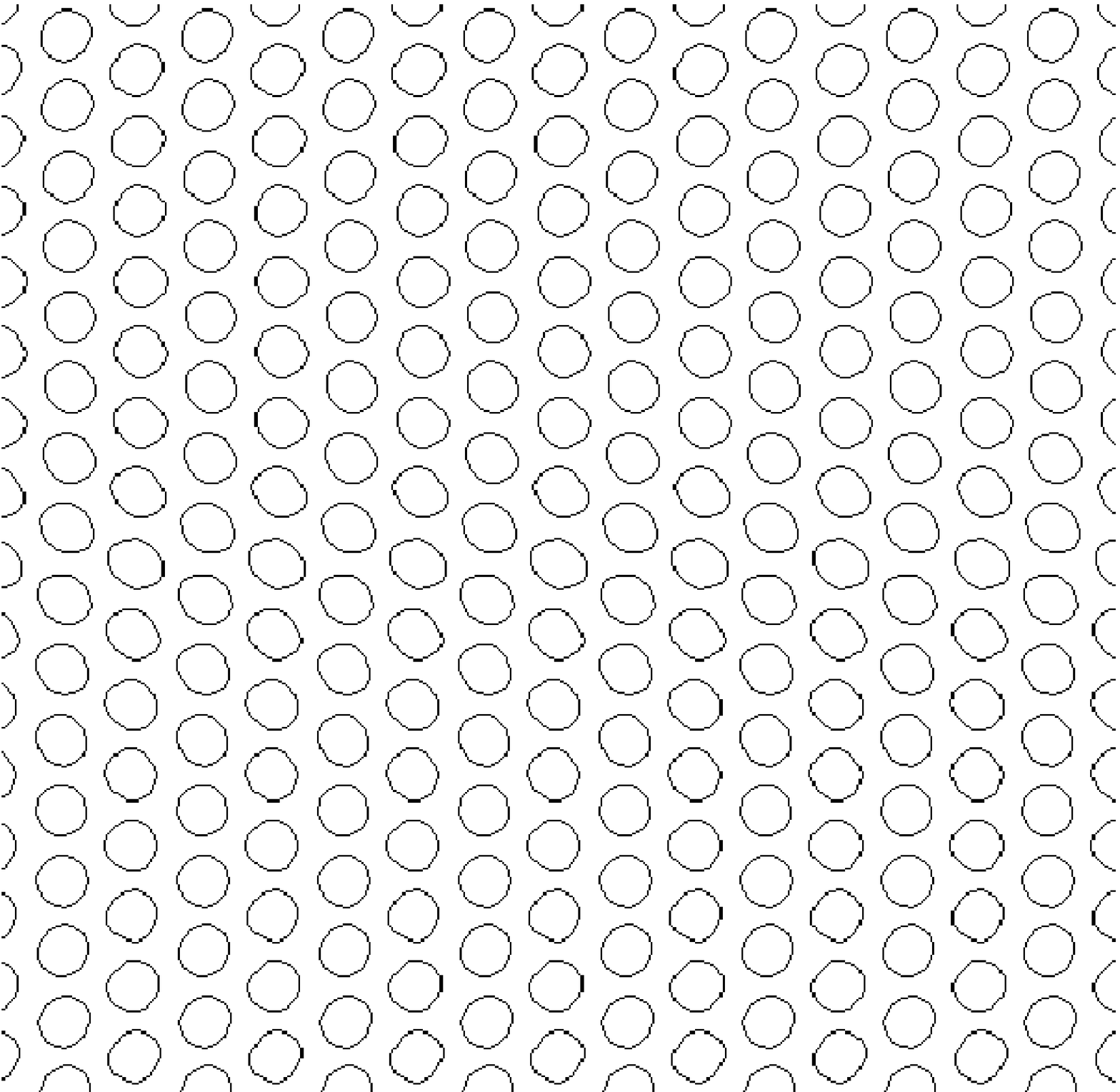}}
  }
\caption{A real-space illustration of the stable oscillations of the
perturbation mode for $R=0.0355$, $k=1.02$, $\alpha=0.1$, $\beta=0$, 
$\gamma=6$, $g_{2}=0$ at times $t=0$, $t=50$, $t=110$, $t=170$.
Indicated in the figure are the zero level contours of the solution.
}
\label{fig:oschex}
\end{figure}

When the wavenumber $k$ or the control parameter $R$ is increased beyond
the line marked by
diamonds in Fig.\ref{fig:a.g.vary.gamma}b the
modulated hexagons become unstable and the regular pattern breaks down
leading to a temporally and spatially
chaotic state.
Typical snapshots are displayed in Fig.\ref{fig:disorosc}a,b. They suggest a strong
signature of a hexagonal symmetry and a predominant
orientation of the disordered pattern which rotates in time. This is brought out
more clearly in the power spectrum of Fig.\ref{fig:disorosc}a pictured in
 Fig.\ref{fig:disorosc}c, which clearly shows six dominant peaks 
(see also Fig.\ref{fig:fourxtric} below). 
Its temporal evolution is depicted in Fig.\ref{fig:fourxt}a as a space-time
diagram of the radially integrated power spectrum 
Fig.\ref{fig:disoroscfour}c with time increasing upward. 
Since $\psi$ is real, only the range $-90^o$ to $90^o$ is shown.
The rotation of the spectrum is seen to be surprisingly pronounced and steady over 
long periods of time, which are intermittently 
interrupted by periods during which the peaks in the spectrum are less pronounced.  
The space-time autocorrelation function of this state as obtained from a run of
duration $T_{max}=278,880$ is shown in Fig.\ref{fig:fourxt}b. The azimuthal decay of the
correlation function is seen to be quite slow.
We do not expect that the sixfold symmetry of the power spectrum
persists as the system size is increased. In fact, the highest peaks in the 
power spectrum decrease by roughly a factor of two if the linear dimensions of the
system are doubled.  However, Fig.\ref{fig:fourxt}a,b indicates that
the disordered state at least {\it locally} exhibits a structure with hexagonal
symmetry, which rotates in time. 

\setlength{\htw}{0.5\columnwidth}
\begin{figure}[h]
 \centerline{
\epsfxsize=0.45\htw{\epsfbox{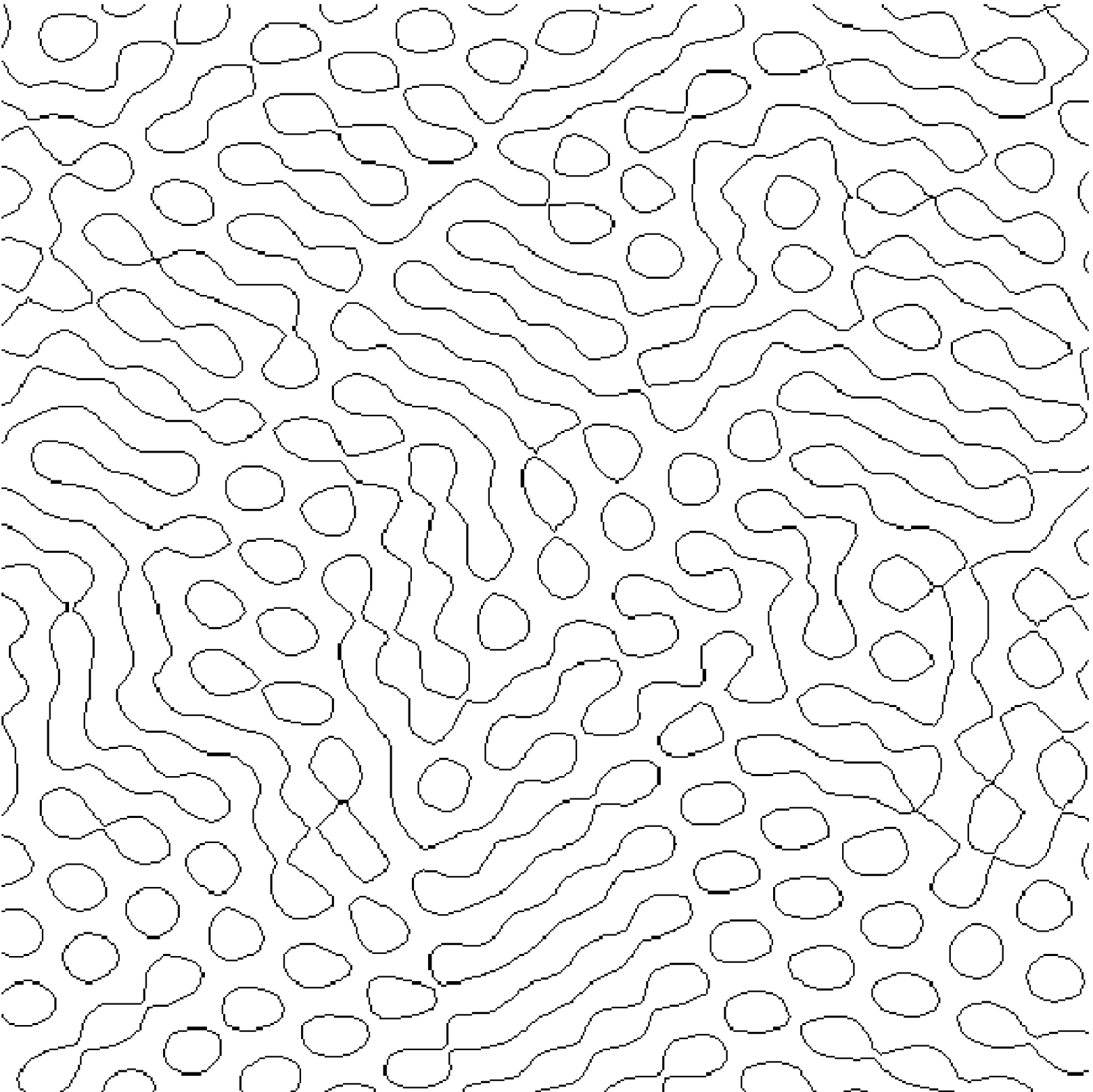}}
\epsfxsize=0.45\htw{\epsfbox{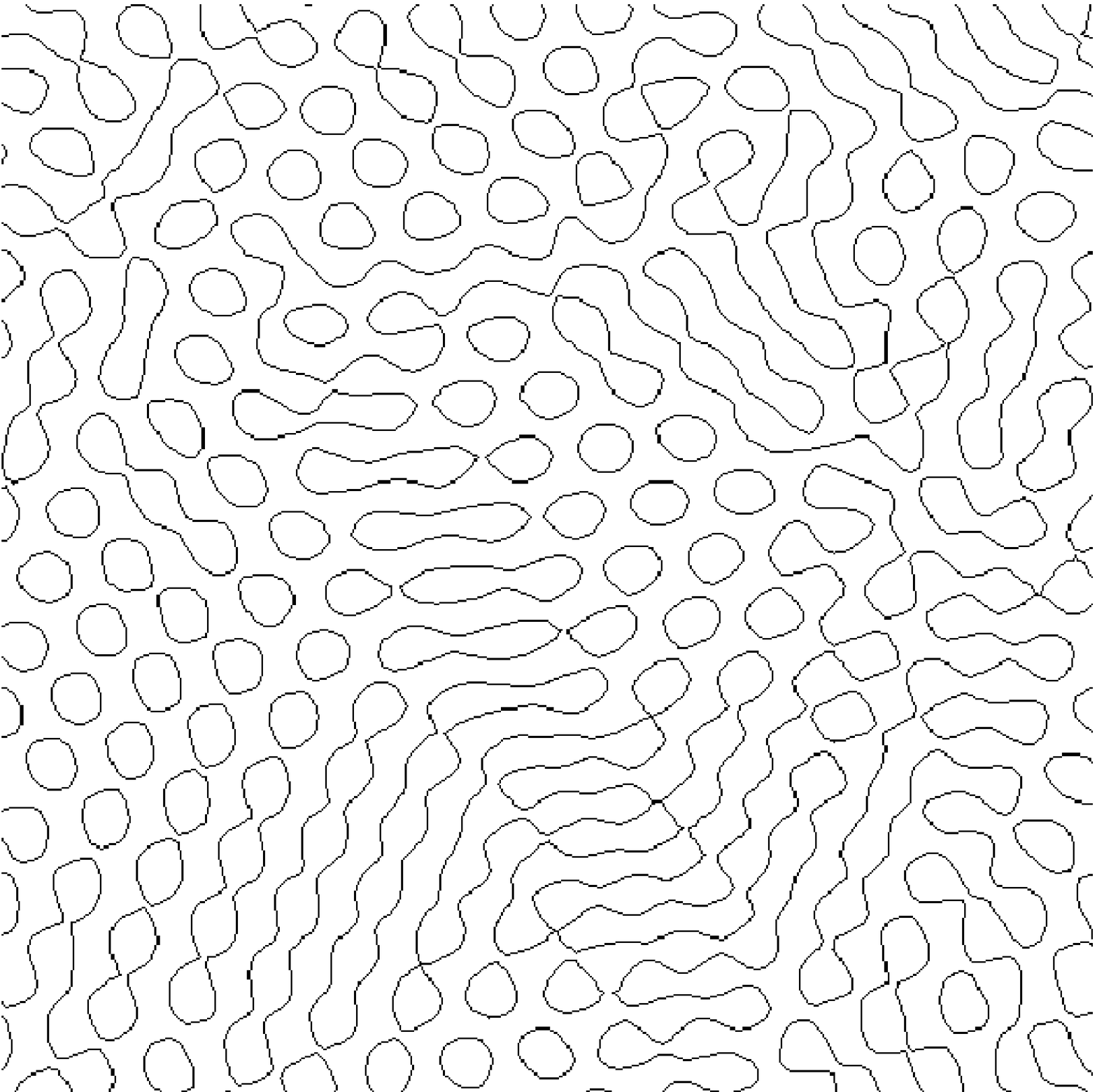}}
}
\centerline{
\epsfxsize=0.6\htw{\epsfbox{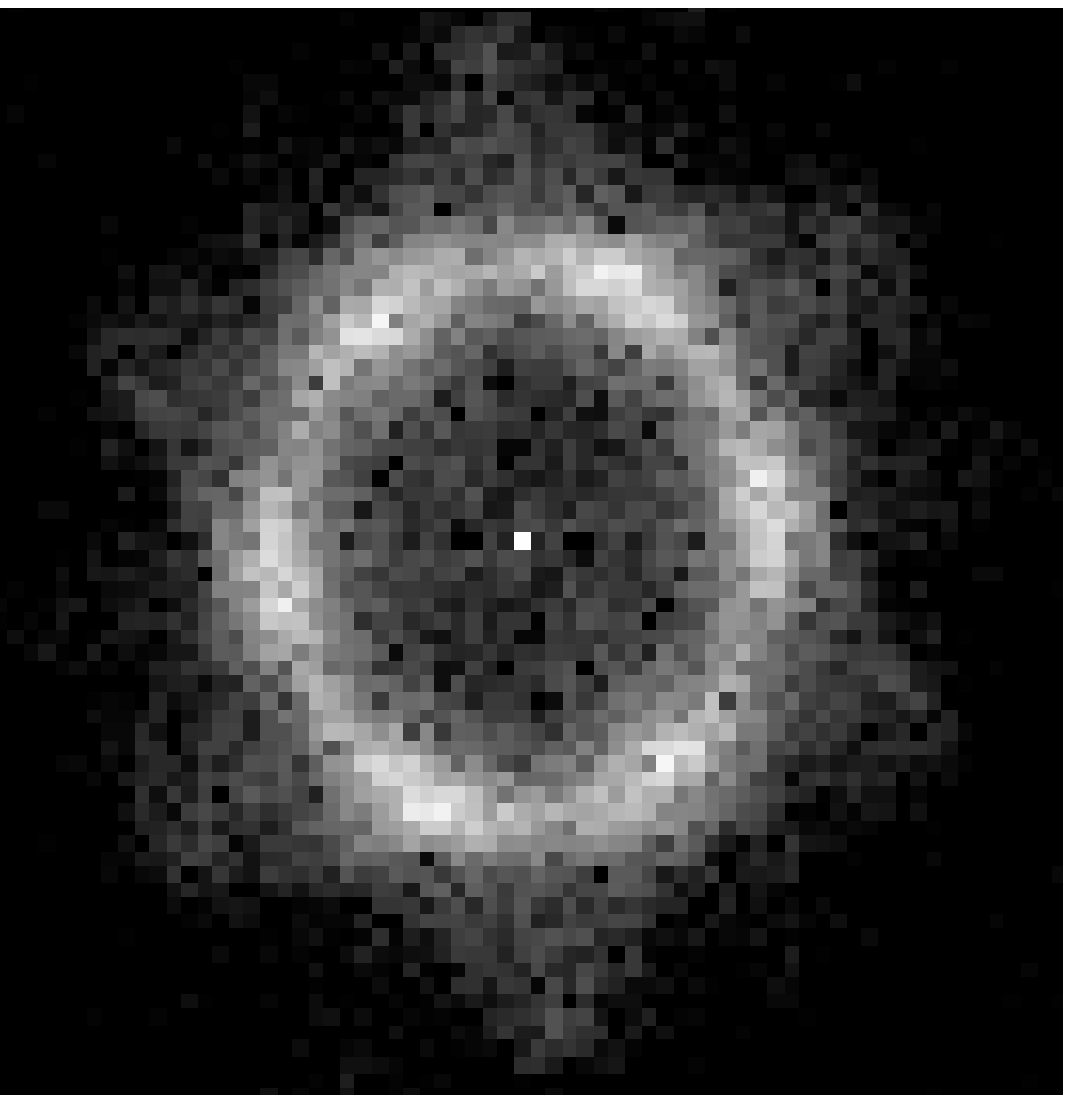}}
}
\caption{
{\it a)} Zero-level contours in real space for time $t=118580$.
{\it b)} Zero-level contours at a time when the power spectrum has
rotated by about $30^o$.
{\it c)} Power spectrum on a logarithmic scale of pattern at time $t=118580$ for
$\alpha = 0.1$, $\gamma = 6$, $R=0.036$, $k= 1.02$. Note the residual sixfold symmetry.
}
\label{fig:disoroscfour}
\label{fig:disorosc}
\end{figure}

\setlength{\htw}{0.5\columnwidth}
\begin{figure}[h]
 \centerline{\epsfxsize=.9\htw{\epsfbox{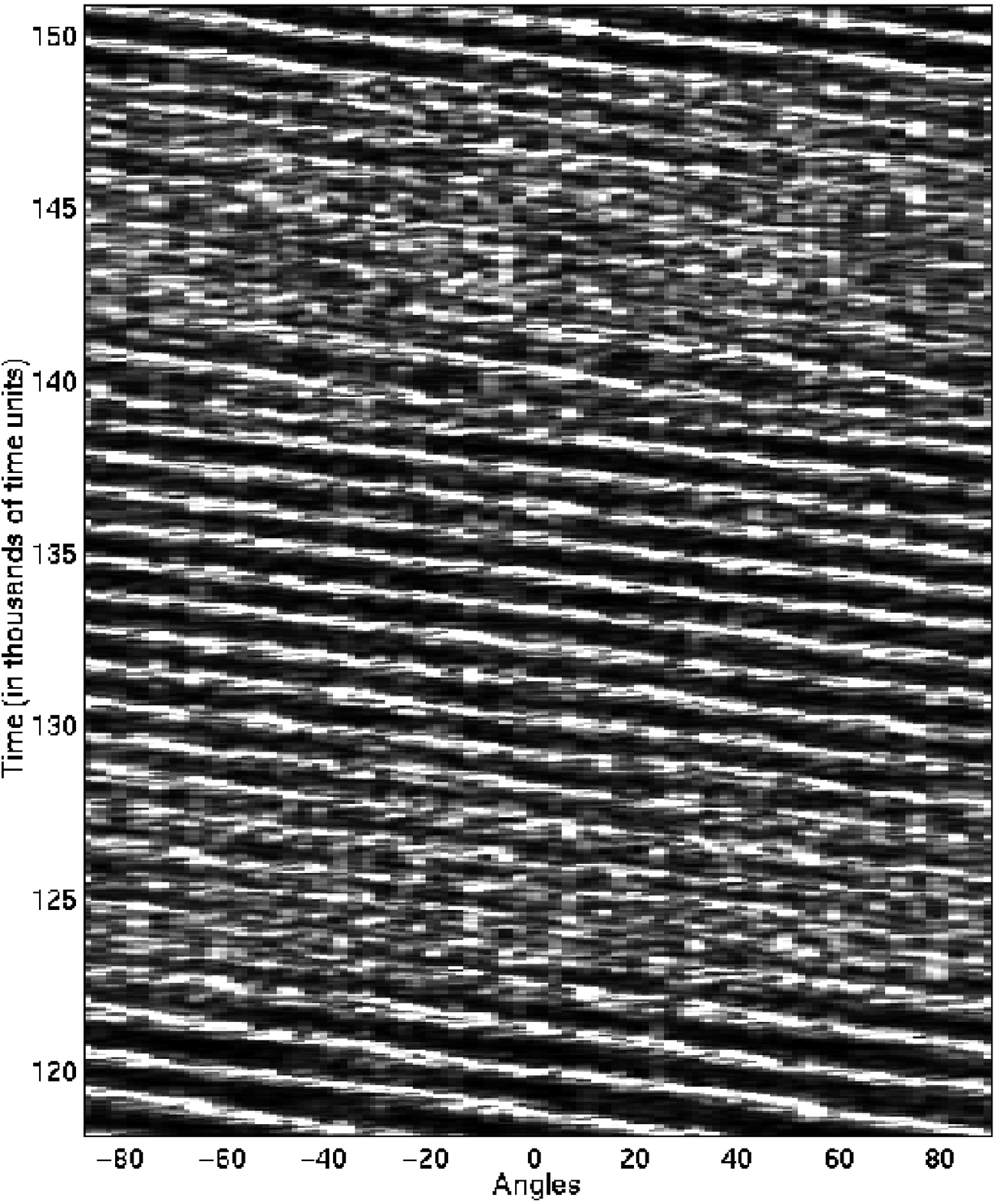}}
             \epsfxsize=.91\htw{\epsfbox{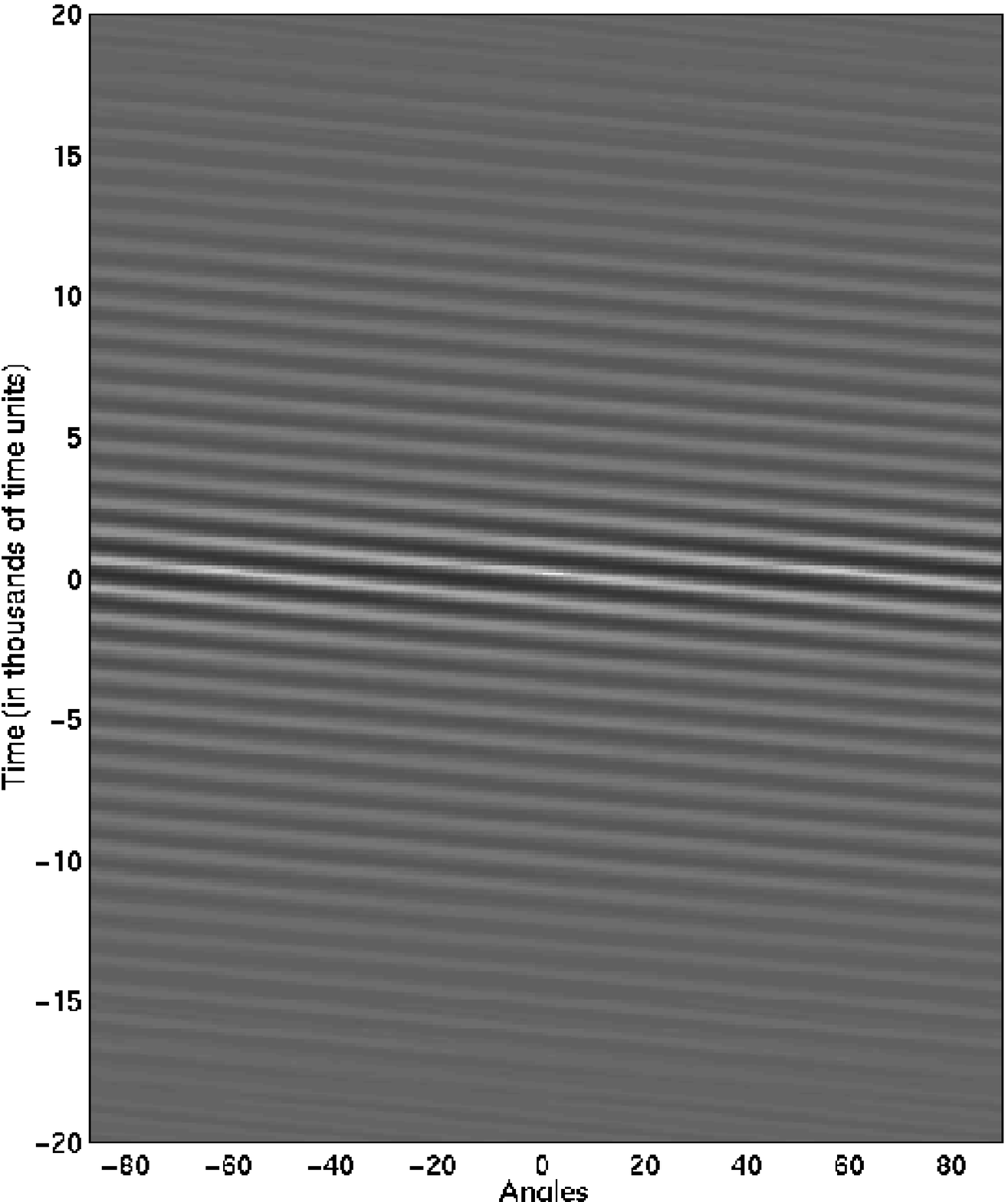}}
	     }
\caption{Temporal evolution of radially integrated power spectrum and space-time
correlation thereof in the disordered domain $\alpha=0.1$, $\gamma=6$, $R=0.036$,
$k=1.02$ (cf. Fig.\protect\ref{fig:detail}).
}
\label{fig:fourxt}
\end{figure}

The modulated hexagons and the disordered hexagons
shown in Fig.\ref{fig:oschex} and Fig.\ref{fig:disorosc} are obtained 
in the minimal model (\ref{eq:mshe}) 
and arise from the short-wave instability connected with the translation mode.
Very similar results are also obtained in the more general model (\ref{eq:mshe2})
where they can also arise from the short-wave Hopf mode. 
For the nonlinear evolution the long-wave behavior
of the branch of destabilizing modes is apparently not relevant.

An interesting aspect of the disordered state of Fig.\ref{fig:disorosc}
is that it exists stably
for the same value of the control parameter as the steady hexagons and the 
modulated hexagons. This coexistence of
ordered and disordered states is reminiscent of the situation in Rayleigh-B\'enard 
convection with small Prandtl number where the state of spiral-defect chaos coexists 
with that of straight parallel rolls \cite{MoBo93}. 
 
It is less surprising to find complex dynamics in parameter regimes in which no
stable steady hexagons exist. Such a case is shown in Fig.\ref{fig:disorgap}. 
The parameters are as in Fig.\ref{fig:a.g.vary.gamma}c with $R$ (marked by a line in
Fig.\ref{fig:a.g.vary.gamma}c) chosen in the gap between the stable steady hexagons
and the transition line to unmodulated oscillating hexagons. In this case 
the disorder is sufficiently strong that the
power spectrum is essentially isotropic for the system sizes investigated.

\setlength{\htw}{0.5\columnwidth}
\begin{figure}[h]
 \centerline{\epsfxsize=0.4\htw{\epsfbox{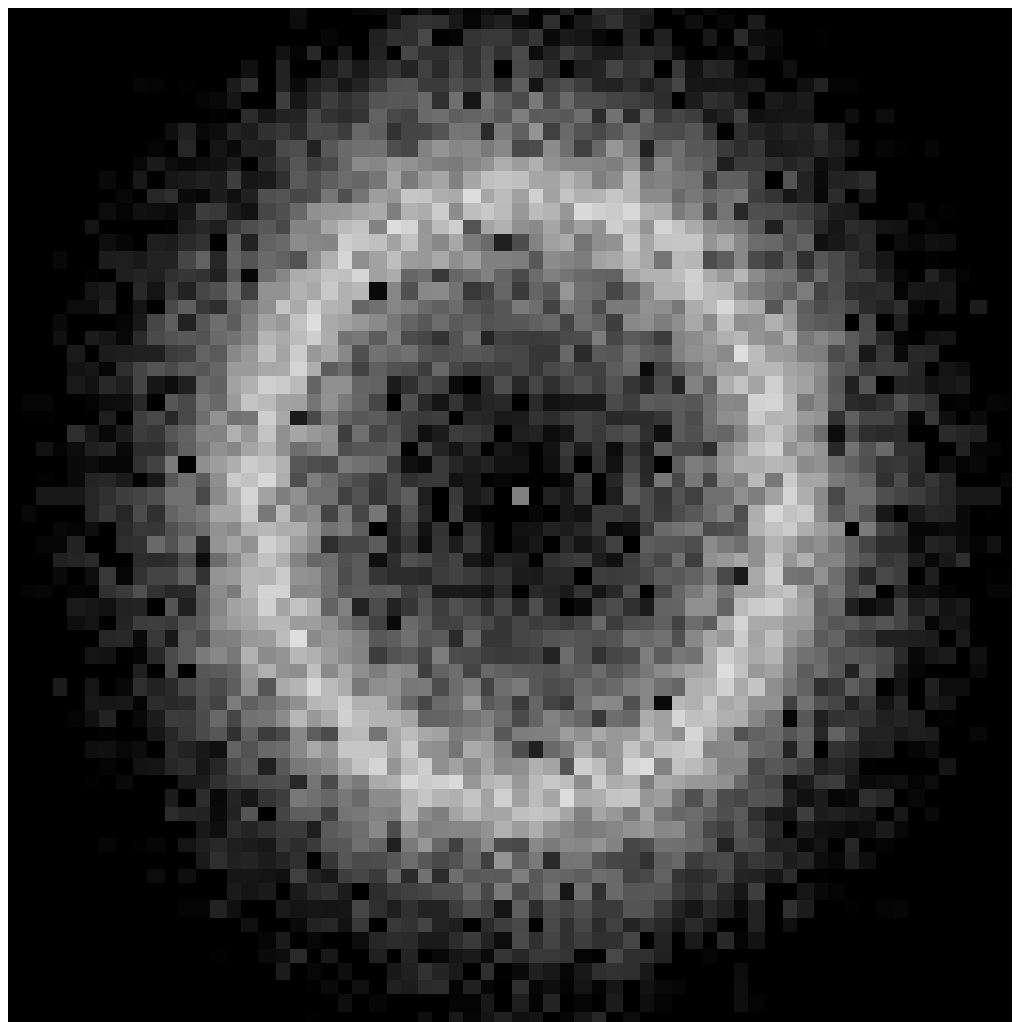}}
             \epsfxsize=0.41\htw{\epsfbox{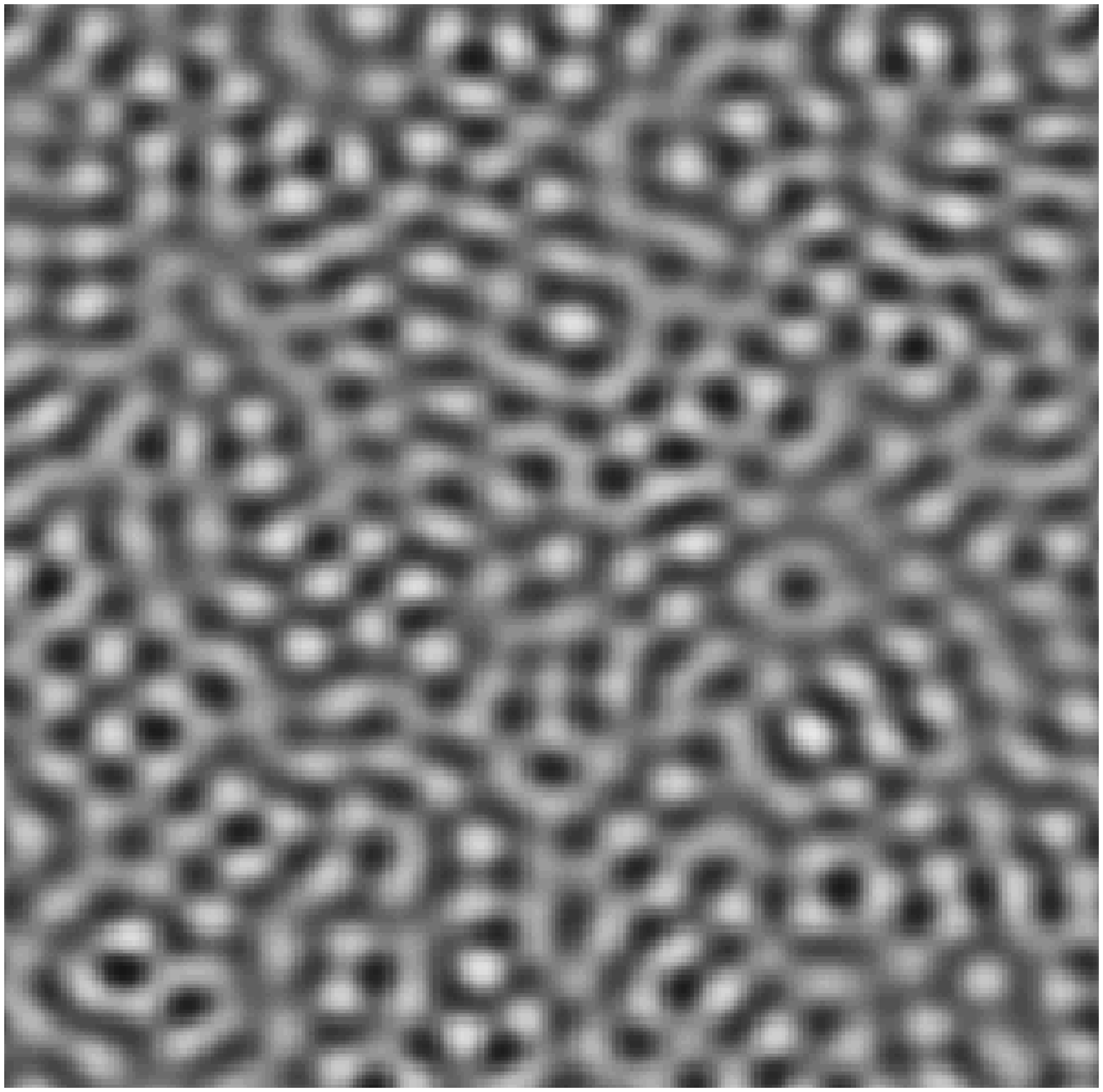}}}
\caption{
Power spectrum on a logarithmic scale and zero-level contour lines of real space
of disordered state at $\alpha=0.1$, $\gamma=14$, $R=0.052$, at
time $t=22400$ in the
minimal model (cf. Fig.\protect\ref{fig:a.g.vary.gamma}c).
}
\label{fig:disorgap}
\end{figure}
 
From an experimental point of view an important question is whether the
interesting dynamic states identified here 
are actually accessible even if the initial conditions cannot be
prepared carefully. After all, they occur in parameter regimes in which the
system could also evolve into a stripe pattern. 
We have therefore performed a few simulations starting from 
random initial conditions. For the parameters in the stability gap of 
Fig.\ref{fig:a.g.vary.gamma}c the system evolves directly to a disordered state
like that shown in Fig.\ref{fig:disorgap}. 
Fig.\ref{fig:fourxtric} shows the space-time diagram for a 
simulation starting from small random initial conditions 
 for parameter values for which steady and modulated hexagons coexist stably with
the disordered state (cf. Fig.\ref{fig:disorosc}). Clearly, the spectrum is 
initially essentially isotropic (bottom panel), 
reflecting the random initial conditions, and then evolves into
a more ordered spectrum (top panel) corresponding to the rotating state discussed above. 
These runs suggest that the states found here should be experimentally accessible.
Of course, so far no information is available to what extent the Swift-Hohenberg 
models discussed here describe a concrete physical system and 
whether the parameter regimes in which the dynamical states arise
are accessible experimentally. 

\begin{figure}[h]
 \centerline{
\epsfxsize=0.8\htw{\epsfbox{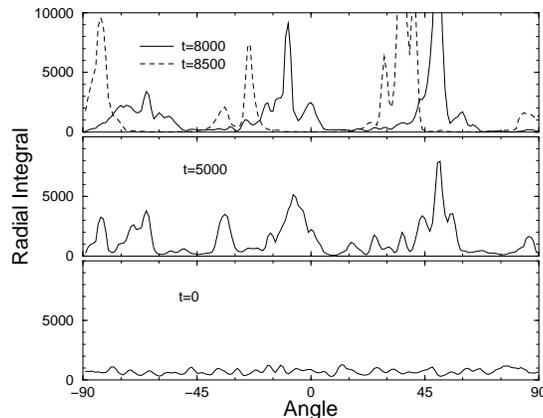}}
	     }
 \caption{Temporal evolution of radially integrated power spectrum 
starting from random initial conditions for  $\alpha=0.1$, $\gamma=6$, $R=0.03$
(cf. Fig.\protect\ref{fig:detail}).
}
\label{fig:fourxtric}
\end{figure}

%%%%%%%%%%%%%%%%%%%%%%%%%%%%%%%%%%%%%%%%%%%%%%%%%%%%%%%%%%%%%%%%%%%%%%%%%%%%%%

\section{Conclusion}

Motivated by the K\"uppers-Lortz instability of roll patterns in the 
presence of rotation we have investigated the effect of rotation on 
the stability of hexagonal patterns. To get a first impression we
have studied a minimal and a more general
Swift-Hohenberg-type order-parameter model. In a
linear stability
analysis of the weakly nonlinear hexagon pattern we focussed on
instabilities
that lead out of the space spanned by the modes corresponding to the 
hexagons themselves. Long- and short-wave instabilities were found, which can be
steady or oscillatory.
Immediately at onset the wavenumber band is limited by
an instability arising from the translation mode. 
For larger rotation rates it may
be oscillatory. In contrast to the case of roll patterns in the presence of rotation, which
for sufficiently large rotation rates can become unstable
at all wavenumbers immediately at onset
due to the K\"uppers-Lortz instability, no such parameter range is found for
 the hexagons.

Further above threshold the instabilities typically become short-wave
and oscillatory. They can lead to periodically oscillating spatially modulated hexagons or
to a disordered state characterized by cellular and stripe-like elements. Locally the
structure can exhibit a surprisingly regular rotation of the orientation of its 
residual hexagonal symmetry. It is worth comparing this behavior with that arising
in the K\"uppers-Lortz regime of systems with up-down symmetry. While in numerical
simulations of an order-parameter model a very pronounced switching between the three
orientations of the stripe pattern was found \cite{CrMe94}, in experiments also 
a behavior was observed that is more akin to the rotation of the spectrum found here
\cite{HuEc95}. Particularly striking is the fact
that the disordered state found here can arise for the same control parameter value as 
the periodically oscillating modulated hexagons and the steady hexagons.
This apparent stable coexistence of the spatio-temporally
chaotic state with an ordered state is reminiscent of the 
spiral-defect chaos found in Rayleigh-B\'enard convection, which coexists with the
straight-roll state \cite{CaEg97}. An interesting question is whether propagating fronts
separating the ordered from the disordered state can be found as has been
in those experiments.

In investigations of the stability of hexagons with broken chiral symmetry within the
framework of coupled Ginzburg-Landau equations
the same type of long- and short-wave, steady and oscillatory instabilities
 have been identified recently \cite{EcRi99}. There also a regime was found in which
the instability does not lead to an ordered state although such a state is stable for
the same value of the control parameter. In the Ginzburg-Landau equations, however, 
that
state cannot be described appropriately. They break the isotropy of the system and
can only describe perturbations that make a small angle with respect to the hexagon
pattern. Since the disordered state that was found in the Swift-Hohenberg equations
investigated here exhibits an almost isotropic Fourier spectrum 
(cf. Figs.\ref{fig:disoroscfour},\ref{fig:disorgap}), it
leads to the excitation of ever higher Fourier modes in the numerical
simulation of the Ginzburg-Landau equations. 
This renders them useless for the investigation
of the chaotic state.  

In order to make concrete predictions for experimental systems the 
order-parameter
equations have to be derived for specific physical systems. 
Situations of interest are Marangoni convection and Rayleigh-B\'enard 
convection with broken Boussinesq symmetry. For Marangoni convection 
with poorly conducting boundaries  \cite{Kn90,GeSi81} the effect of rotation has  only
been included recently in the derivation of long-wave equations \cite{MaSaunpub}. The case
of Rayleigh-B\'enard convection has been treated to some extent.
For poorly conducting boundaries a systematic long-wave expansion has 
been performed \cite{Co98}. By imposing asymmetric boundary conditions 
(but neglecting other non-Boussinesq effects) quadratic terms in the order parameter 
equation were obtained, including the crucial term involving $\gamma$ that 
breaks the chiral symmetry. In that work the analysis focused on 
square rather than hexagonal patterns \cite{Co98}. 

If the pattern arises
first at a finite (i.e. not small) wavenumber an equation of the form
(\ref{eq:mshe}) or (\ref{eq:mshe2}) can be obtained in an approximation in which the 
kernel of the
nonlinear integral term in Fourier space is approximated by local 
terms
in real space \cite{Ha83a,BeFr98,BeFr99}. For Boussinesq
Rayleigh-B\'enard convection
with and without rotation it has been found that such an 
approximation can yield good agreement with weakly nonlinear 
amplitude equations while still preserving the isotropy of the system
\cite{NeFr93}. 
The non-Boussinesq case with rotation has not been treated yet.

We gratefully acknowledge helpful discussions with  B. Echebarria, A. Golovin,
A. Mancho, W. Pesch, and M. Silber.
The computations were done with a modification of a code by G. D. 
Granzow.
This work was supported by D.O.E. Grant DE-FG02-G2ER14303 and NASA 
Grant NAG3-2113.

\bibliography{journal}
\bibliographystyle{prsty}
%%%%%%%%%%%%%%%%%%%%%%%%%%%%%%%%%%%%%%%%%%%%%%%%%%%%%%%%%%%%%%%%%%%%%%%%%%%%%%%%%%%%%%%%%%%%%%%%%

\end{document}